\documentclass[superscriptaddress,amsmath,amssymb,nofootinbib,eqsecnum,a4paper,secnumarabic,11pt]{revtex4}         
\pdfoutput = 1
\usepackage[pdftex]{graphicx}
\usepackage{hyperref}     
\usepackage{grffile} 
\usepackage{slashed,color,braket}

\newcommand{\VEV}[1]{\langle #1 \rangle}
\usepackage{titlesec}     
\newcommand*{\justifyheading}{\raggedright}
\titleformat{\chapter}[display]
  {\normalfont\huge\bfseries\justifyheading}{\chaptertitlename\ \thechapter}
  {20pt}{\Huge}
\titleformat{\section}
  {\normalfont\Large\bfseries\justifyheading}{\thesection}{1em}{}
\titleformat{\subsection}
  {\normalfont\large\bfseries\justifyheading}{\thesubsection}{1em}{}
\titleformat{\subsubsection}
  {\normalfont\bfseries\justifyheading}{\thesubsubsection}{1em}{}
\usepackage[english]{babel}

\begin{document}
\begin{flushright}
{\tt 
IPMU19-0142
}
\end{flushright}
\title{Maximum value of the spin-independent cross section in the 2HDM+a}
\author{Tomohiro Abe}
\email{abetomo@kmi.nagoya-u.ac.jp}
\affiliation{
  Institute for Advanced Research, Nagoya University,
  Furo-cho Chikusa-ku, Nagoya, 464-8602 Japan
}
\affiliation{
  Kobayashi-Maskawa Institute for the Origin of Particles and the
  Universe, Nagoya University,
  Furo-cho Chikusa-ku, Nagoya, 464-8602 Japan
}

\author{Motoko Fujiwara}
\email{motoko@eken.phys.nagoya-u.ac.jp}
\affiliation{
  Department of Physics, Nagoya University, Furo-cho Chikusa-ku, Nagoya, 464-8602 Japan
}

\author{Junji Hisano}
\email{hisano@eken.phys.nagoya-u.ac.jp}
\affiliation{
  Kobayashi-Maskawa Institute for the Origin of Particles and the
  Universe, Nagoya University,
  Furo-cho Chikusa-ku, Nagoya, 464-8602 Japan
}
\affiliation{
  Department of Physics, Nagoya University, Furo-cho Chikusa-ku, Nagoya, 464-8602 Japan
}
\affiliation{
  Kavli IPMU (WPI), UTIAS, University of Tokyo, Kashiwa, 277-8584, Japan
}
\author{Yutaro Shoji}
\email{yshoji@kmi.nagoya-u.ac.jp}
\affiliation{
  Kobayashi-Maskawa Institute for the Origin of Particles and the
  Universe, Nagoya University,
  Furo-cho Chikusa-ku, Nagoya, 464-8602 Japan
}

\begin{abstract}
We investigate the maximum value of the spin-independent cross section ($\sigma_\text{SI}$)
in a dark matter (DM) model called the two-Higgs doublet model + a (2HDM+a). 
This model can explain the measured value of the DM energy density by the freeze-out mechanism.
Also, $\sigma_\text{SI}$ is suppressed by the momentum transfer at the tree level,
and loop diagrams give the leading contribution to it.
The model prediction of $\sigma_\text{SI}$ highly depends on
values of $c_1$ and $c_2$ that are the quartic couplings between the gauge singlet CP-odd state ($a_0$) and Higgs doublet fields ($H_1$ and $H_2$), $c_1 a_0^2 H_1^\dagger H_1$ and $c_2 a_0^2 H_2^\dagger H_2$.
We discuss the upper and lower bounds on $c_1$ and $c_2$ by studying 
the stability of the electroweak vacuum,
the condition for the potential bounded from the below,
and the perturbative unitarity.
We find that the condition for the stability of the electroweak vacuum gives upper bounds on 
$c_1$ and $c_2$.
The condition for the potential to be bounded from below gives lower bounds on $c_1$ and $c_2$.
It also constrains the mixing angle between the two CP-odd states.
The perturbative unitarity bound gives the upper bound on the Yukawa coupling between the dark matter and $a_0$ and the quartic coupling of $a_0$.
Under these theoretical constraints, we find that the maximum value of the $\sigma_\text{SI}$ is $\sim 5\times 10^{-47}$~cm$^2$ for
$m_A = $~600~GeV, and the LZ and XENONnT experiments can see the DM signal predicted in this model near future.
\end{abstract}

\maketitle

\section{Introduction}

One of the great achievements in Cosmology is the precise determination of the energy density of the dark matter (DM) by the Planck collaboration, $\Omega h^2 = 0.120\pm 0.001$~\cite{1807.06209}.
The measured value is explained successfully by DM models that use the freeze-out mechanism~\cite{Lee:1977ua}, which have been widely studied 
for a long time.
Those models generally predict non-zero DM-nucleon scattering cross section and have been searched by
the direct detection experiments, such as the Xenon1T experiment~\cite{1805.12562}.
However, no significant signals have been observed until now, and
the null results set upper bound on the DM-nucleon scattering cross section.
The latest result by the Xenon1T experiment gives a severe constraint on DM models.

If a DM particle is a gauge singlet fermion, $\chi$, and couples to a scalar mediator, $a_0$, with pseudo-scalar type
interaction, $\bar{\chi} i \gamma_5 \chi a_0$,
then it is possible to avoid this strong constraint from the Xenon1T experiment 
while keeping the success of the freeze-out mechanism~\cite{1609.09079, 1612.06462}.
The two-Higgs doublet model + a (2HDM+a)~\cite{1404.3716} is one of 
the models that realize this idea.\footnote{
Other realizations are discussed in, for example, Refs.~\cite{1408.4929, 1701.04131}. 
Another mechanism to avoid the constraint from direct detection experiments 
is studied in Ref.~\cite{1708.02253}.
}
In addition to the introduction of the DM and the mediator, the Higgs sector is extended into the two-Higgs doublet model.
The CP invariance is assumed in the dark sector and the scalar sector.
Then, the dark sector and the visible sector can interact 
through the mixing between $a_0$ and the CP-odd scalar ($A_0$) in the two-Higgs doublet sector.
The model predicts rich phenomenology~\cite{1509.01110, 1611.04593, 1701.07427, 1705.09670, 1711.02110, 1712.03874, 1803.01574, 1804.02120}, and it is summarized in Ref.~\cite{1810.09420}.

The 2HDM+a predicts non-zero spin-independent DM-nucleon scattering cross section ($\sigma_\text{SI}$) at loop level~\cite{1404.3716,1711.02110, 1803.01574, 1804.02120, 1810.01039}.
In particular, it was shown that if $c_2$, which is a quartic coupling between $a_0$ and a Higgs doublet field $H_2$, is large enough, the model can be tested
at the forthcoming direct detection experiments~\cite{1810.01039}.
However, such a large coupling causes theoretical problems.
If the coupling takes large negative value, the potential can be unbounded from the below.
If the coupling is very large, it hits a Landau pole near the electroweak scale
and the model loses predictability.

In this paper, we study the constraint on the scalar potential from
the boundedness of the scalar potential, 
the stability of the electroweak vacuum,
and perturbative unitarity.
Using these constraints, we investigate the upper and the lower bounds on the scalar quartic couplings,
and discuss the maximum value of $\sigma_\text{SI}$.
We show that the maximum value of $\sigma_\text{SI}$ is below
the current constraint from the Xenon1T experiment 
and above the prospect of the LZ experiment~\cite{1802.06039} and the XENONnT experiment~\cite{Aprile:2015uzo}.

The rest of the paper is organized as follows.
In Sec.~\ref{sec:model}, we briefly describe the 2HDM+a.
In Sec.~\ref{sec:constraints}, we investigate theoretical constraints on the model parameters.
Conditions for the electroweak vacuum as the global minimum of the scalar potential, 
for the potential to be bounded from below, 
and for perturbative unitarity for the quartic couplings in the scalar potential
are discussed. 
These conditions are used to find the upper and the lower bounds on $c_1$ and $c_2$
and the upper bound on the mixing angle between the two CP-odd states.
In Sec.~\ref{sec:main}, we scan the model parameter space 
and find the maximum value of $\sigma_\text{SI}$ under the constraint
discussed in Sec.~\ref{sec:constraints}.
Section~\ref{sec:summary} is devoted to our conclusion.

\section{Model}\label{sec:model}
The model contains a gauge singlet Majorana fermion $\chi$ as a DM candidate 
and a CP-odd gauge singlet scalar $a_0$ as a mediator.
The standard model (SM) Higgs sector is extended into the two-Higgs doublet model. 
We assume CP invariance both in the dark sector and in the scalar sector.
This assumption guarantees that the Yukawa interaction between $\chi$ and $a_0$ is always pseudo-scalar interaction.
The Lagrangian is given by
\begin{align}
 {\cal L}
=& 
\frac{1}{2}\bar{\chi}  \left( i \slashed{\partial} - m_{\rm DM} \right) \chi
- \frac{g_\chi}{2}\bar{\chi} i \gamma^5 \chi a_0
\nonumber\\
&
+ \frac{1}{2} \partial_\mu a_0 \partial^\mu a_0 
+ D_\mu H_1^\dagger D^\mu H_1
+ D_\mu H_2^\dagger D^\mu H_2
- V_\text{scalar}
\nonumber\\
&
+ \text{(terms with the SM fermions and gauge bosons)}
,
\end{align}
where
\begin{align}
 V_\text{scalar}=&
m_1^2 H_1^{\dagger} H_1
+
m_2^2 H_2^{\dagger} H_2
-
m_3^2
\left(
H_1^{\dagger} H_2
+
(h.c.)
\right)
\nonumber\\
&
+ \frac{1}{2} \lambda_1 (H_1^{\dagger} H_1)^2
+ \frac{1}{2} \lambda_2 (H_2^{\dagger} H_2)^2
+ \lambda_3 (H_1^{\dagger} H_1) (H_2^{\dagger} H_2)
+ \lambda_4 (H_1^{\dagger} H_2) (H_2^{\dagger} H_1)
\nonumber\\
&
+
\frac{1}{2}
\lambda_5
\left(
 (H_1^{\dagger} H_2)^2 
+
(h.c.)
\right)
\nonumber\\
&
+\frac{1}{2} m_{a_0}^2 a_0^2
+\frac{\lambda_a}{4} a_0^4
+ \kappa \left( i a_0  H_1^\dagger H_2 + (h.c.) \right)
+ c_1 a_0^2 H_1^\dagger H_1
+ c_2 a_0^2 H_2^\dagger H_2
.
\end{align}
Since we assume the CP invariant scalar potential,
all the couplings in the potential are real.
In this paper, we assume that the thermal relic abundance of $\chi$ explains the measured value of the DM energy density~\cite{1807.06209}, and $g_\chi$ is fixed to realized it for a given parameter set by the freeze-out mechanism.

We impose the condition that the potential has the electroweak vacuum,
\begin{align}
\VEV{a_0}
=0
,\
\VEV{H_1}
=&
\left(
\begin{matrix}
 0  \\  \frac{1}{\sqrt{2}} v_1
\end{matrix}
\right)
,\
\VEV{H_2}
=
\left(
\begin{matrix}
 0  \\  \frac{1}{\sqrt{2}} v_2
\end{matrix}
\right)
.
\end{align}
This electroweak vacuum is realized if $m_1^2$ and $m_2^2$ satisfy the following relations.
\begin{align}
 m_1^2 =&
 - \frac{v_1^2 \lambda_1 + v_2^2 \lambda_{345}}{2}
+ m_3^2 \frac{v_2}{v_1}
,\\
 m_2^2 =&
 - \frac{v_2^2 \lambda_2 + v_1^2 \lambda_{345}}{2}
+ m_3^2 \frac{v_1}{v_2}
,
\end{align}
where $\lambda_{345} = \lambda_3 + \lambda_4 + \lambda_5$.
In the following, we assume $m_1^2$ and $m_2^2$ always satisfy these relations.
It is also important that $a_0$ does not develop vacuum expectation value. 
Otherwise, the scalar-type Yukawa interaction
is induced in the dark sector due to the scalar and pseudo-scalar mixing, and the model 
is strongly constrained from the direct detection experiments.

After the electroweak symmetry breaking, 
there are two CP-even scalars ($h$ and $H$), 
two CP-odd scalars ($a$ and $A$), 
a pair of charged scalars ($H^\pm$),
and three would-be Nambu-Goldstone bosons that are eaten by $W^\pm$ and $Z$. The physical masses for $h$, $H$, $a$, $A$, and $H^\pm$ are denoted to $m_h$, $m_H$, $m_a$, $m_A$, and $m_{H^\pm}$, respectively. 
The two CP-even scalars are mixtures of the CP-even neutral components in $H_1$ and $H_2$,
and its mixing angle is denoted by $\alpha$.
Similarly, the two CP-odd scalars are mixtures of the CP-odd neutral components in $H_1$ and $H_2$ and also $a_0$. Its mixing angle is denoted by $\theta$.

We introduce the following notations for later convenience,
\begin{align}
 t_\beta =& \tan\beta = \frac{v_2}{v_1},\quad
 s_\beta = \sin\beta, \quad
 c_\beta = \cos\beta,\\
 v =& \sqrt{v_1^2 + v_2^2},\\
 M^2 =& \frac{v_1^2 + v_2^2}{v_1 v_2} m_3^2. \label{eq:M2-vs-m32}
\end{align}

Let us comment on the types of the Yukawa interaction. 
The model is classified into four types based on the Yukawa interaction between the two-Higgs doublet fields and the SM fermions, as in the two-Higgs doublet model with softly broken $Z_2$ symmetry~\cite{Barger:1989fj, Grossman:1994jb, Aoki:2009ha}.
In the following analysis,
we choose the type-I Yukawa interaction where $H_2$ couples to the SM fermions but $H_1$ does not.
The following discussion is independent from the types of the Yukawa interaction
because 
the type dependence is negligible for large $\sigma_\text{SI}$ region of the parameter space
as we showed in our previous work~\cite{1810.01039},
and the purpose of this paper is to find the maximum value of $\sigma_\text{SI}$ for a given parameter set.

We can express $\lambda_i \ (i = 1,2,3,4,5)$, $\kappa$, and $m_{a_0}^2$ by the mixing angles and mass eigenvalues. 
In the followings, we take the mixing angle in the CP-even scalars as
$\alpha = \beta - \pi/2$. This choice predicts the same $hWW$ and the $hZZ$ couplings as in the SM.
We also take $M = m_H = m_A = m_{H^\pm}$. This choice of the mass parameters enhances the
custodial symmetry in the scalar potential, and thus the constraints from the electroweak precision
measurements are automatically satisfied.
With these parameter choices, the parameters of the scalar potential are given by
\begin{align}
 \lambda_1 = \lambda_2 = \lambda_3 =& \frac{m_h^2}{v^2} ,\label{eq:lam123}\\
\lambda_4 = - \lambda_5 = & -\frac{m_A^2  - m_a^2 }{v^2} s_\theta^2 ,\label{eq:lam45}\\
 \kappa =&  -\frac{m_A^2 - m_a^2}{2v} \sin2\theta,\label{eq:kappa}\\
m_{a_0}^2 =&
m_a^2 c_\theta^2 + m_A^2 s_\theta^2 - \frac{c_1 + c_2 t_\beta^2}{1+t_\beta^2} v^2.
\label{eq:ma02}
\end{align}
%
As can be seen, $|\lambda_{1,2,3}| < 1$. 
We can also show that $|\lambda_{4,5}|<1$ with a condition for the boundedness of the scalar potential (Eq.~\eqref{eq:max-theta}) discussed in the next section.

\section{Theoretical constraints on the scalar potential}
\label{sec:constraints}
In this section, we discuss the condition for the electroweak vacuum as the global minimum of the scalar potential, the conditions for the potential to be bounded from below, and the perturbative unitarity for the quartic couplings in the scalar potential. 
These constraints are used to find the upper and the lower bounds on $c_1$, $c_2$, and $\theta$.

\subsection{Vacuum structure}
\label{sec:global-min}
Vacua other than the electroweak vacuum can exist depending on the given parameter sets.
We study the vacuum structure at the tree level and impose the condition that the electroweak vacuum should be the global minimum.
It is not necessary for the electroweak vacuum to be the global minimum if its
lifetime is much longer than the age of our Universe.
However, the lifetime is much shorter than the age of the Universe in most of the parameter space.\footnote{The lifetime is estimated by using \texttt{SimpleBounce}~\cite{1908.10868}.}
Therefore, we adopt the condition to be the global minimum in the current analysis.

\subsubsection{$\VEV{H_1} = \VEV{H_2} = 0$}
In this case, the stationary condition for the scalar potential is given by
\begin{align}
 a_0 \left( m_{a_0}^2 + \lambda_a a_0^2 \right) = 0.
\end{align}
If $m_{a_0}^2 < 0$, we have vacua where $a_0$ develops the vacuum expectation value.
The sign of $m_{a_0}^2$ depends on the values of $c_1$, $c_2$, and $t_\beta$ as in Eq.~(\ref{eq:ma02}).
Since we impose the condition that the electroweak vacuum should be the global minimum,
such vacua should not be deeper than the electroweak vacuum.

At the vacuum with $\langle a_0\rangle\neq0$, the potential energy is given by
\begin{align}
 \left.V_{min.}\right|_{\VEV{H_1} = \VEV{H_2} = 0, \VEV{a_0} \neq 0}
=  -\frac{m_{a_0}^4}{4 \lambda_a}.
\end{align}
This should be larger than the potential energy at the electroweak vacuum,
\begin{align}
 \left.V_{min.}\right|_{\VEV{H_1} \neq 0, \VEV{H_2} \neq 0, \VEV{a_0} = 0}
=
 -\frac{1}{8}
 \left(
  m_h^2 s_{\beta-\alpha}^2 
+ m_H^2 c_{\beta-\alpha}^2
+  \frac{4 (m_{H^\pm}^2 - m_H^2) t_\beta^2}{(1+t_\beta^2)^2} 
\right)
v^2
.
\end{align}
Therefore, we obtain the following condition for $m_{a_0}^2 < 0$,
\begin{align}
\lambda_a
 \left(
  m_h^2 s_{\beta-\alpha}^2 
+ m_H^2 c_{\beta-\alpha}^2
+  \frac{4 (m_{H^\pm}^2 - m_H^2) t_\beta^2}{(1+t_\beta^2)^2} 
\right)
>
 \frac{2 m_{a_0}^4}{v^2}
.
\label{eq:No-a0-vev}
\end{align}
From Eqs.~\eqref{eq:ma02} and \eqref{eq:No-a0-vev},
for $\sin(\beta - \alpha) = 1$ and $M = m_H = m_A = m_{H^\pm}$,
we find
\begin{align}
\frac{c_1 + c_2 t_\beta^2}{1+t_\beta^2} 
<
\sqrt{\frac{\lambda_a m_h^2}{2 v^2}}
+
\frac{m_a^2 c_\theta^2 + m_A^2 s_\theta^2}{v^2}
.
\label{eq:c1c2upperbound}
\end{align}
As a result, we obtain the upper bound on $c_1$ or $c_2$ for a given parameter sets.

\subsubsection{One of $\VEV{H_1}$ and $\VEV{H_2}$ is zero}
We investigate $\VEV{H_1} = 0$ and $\VEV{H_2} \neq 0$.
Without lose of the generality, we can parametrize the vacuum as 
\begin{align}
\VEV{H_2}
=&
\left(
\begin{matrix}
 0 \\ \frac{1}{\sqrt{2}} \sigma_2
\end{matrix}
\right)
.
\end{align}
A stationary condition of this vacuum is given by
\begin{align}
0&= - m_3^2 \sigma_2.
\end{align}
This condition is obtained in both $\VEV{a_0}=0$ and $\VEV{a_0}\neq 0$ cases.
Since $m_3^2 \neq 0$, see Eq.~\eqref{eq:M2-vs-m32}, this condition implies $\sigma_2  = 0$.  
This is contradict to $\VEV{H_2} \neq 0$.
Therefore, we do not have vacua that satisfy $\VEV{H_1} = 0$ and $\VEV{H_2} \neq 0$.

In the same manner, we can show that 
we do not have vacua that satisfy $\VEV{H_1} \neq 0$ and $\VEV{H_2} = 0$.

\subsubsection{$\VEV{H_1} \neq 0 $ and $\VEV{H_2} \neq 0$}
We simplify the analysis as much as possible by using the gauge invariance in the
potential. 
Without lose of the generality, we can parametrize the Higgs fields as
\begin{align}
\VEV{H_1}
=&
\begin{pmatrix}
 0  \\  \frac{1}{\sqrt{2}} \sigma_1
\end{pmatrix}
,
\quad
\VEV{H_2}
=
\begin{pmatrix}
 \pi_2^+  \\  \frac{1}{\sqrt{2}} (\sigma_2 + i \pi^0_{2})
\end{pmatrix}
,
\end{align}
where $\sigma_1$ is positive and $\sigma_2$, $\pi_2^0$, and $\pi_2^+$ are real numbers.
In this case, since the analysis is complicated, we rely on numerical analysis.

\subsection{Conditions for the potential to be bounded from below}
\label{sec:bfb}
The potential should be bounded from below. 
In other words, the potential should be positive for the region where the field values are extremely large.
We find the following seven conditions for the boundedness of the scalar potential.
\begin{align}
 & \lambda_1 >0, \label{eq:BFB_1}\\
 & \lambda_2 >0, \label{eq:BFB_2}\\
 & \lambda_a >0, \label{eq:BFB_3}\\
 & \sqrt{\lambda_1 \lambda_2} + \tilde{\lambda}_3 > 0, \label{eq:BFB_4}\\
 & \sqrt{\frac{\lambda_1 \lambda_a}{2}} + c_1 > 0, \label{eq:BFB_5}\\
 & \sqrt{\frac{\lambda_2 \lambda_a}{2}} + c_2 > 0, \label{eq:BFB_6}\\
& 
\begin{cases}
\sqrt{\lambda_1} c_2 + \sqrt{\lambda_2} c_1 \geq 0,\\
\text{or} \\%
\sqrt{\lambda_1} c_2 + \sqrt{\lambda_2} c_1 <0
\text{ and } 
\frac{\lambda_a \tilde{\lambda}_3}{2} - c_1 c_2 + \sqrt{\left( \frac{\lambda_a \lambda_1}{2} - c_1^2  \right)\left( \frac{\lambda_a \lambda_2}{2} - c_2^2  \right)} > 0.
\end{cases}
\label{eq:BFB_7}
\end{align}
where
\begin{align}
\tilde{\lambda}_3 = \lambda_3+ \min(0, \lambda_4 - |\lambda_5|).
\end{align}
The derivation is given in Appendix~\ref{app:BFB}.\footnote{
The scalar potential discussed in Ref.~\cite{1408.2106} is the same as in this paper, 
but they find that the second condition in Eq.~\eqref{eq:BFB_7} should be applied for $c_1$ or $c_2 < 0$.
The condition given in Ref.~\cite{1612.01309}, which was derived from the result
given in Ref.~\cite{Klimenko:1984qx}, is consistent with our result.
The condition given in Ref.~\cite{1709.08581} is different from ours.
}
As can be seen,
Eqs.~\eqref{eq:BFB_5}, \eqref{eq:BFB_6}, and \eqref{eq:BFB_7} give the lower bounds on $c_1$ and $c_2$.

\begin{figure}[tb]
\includegraphics[width=0.68\hsize]{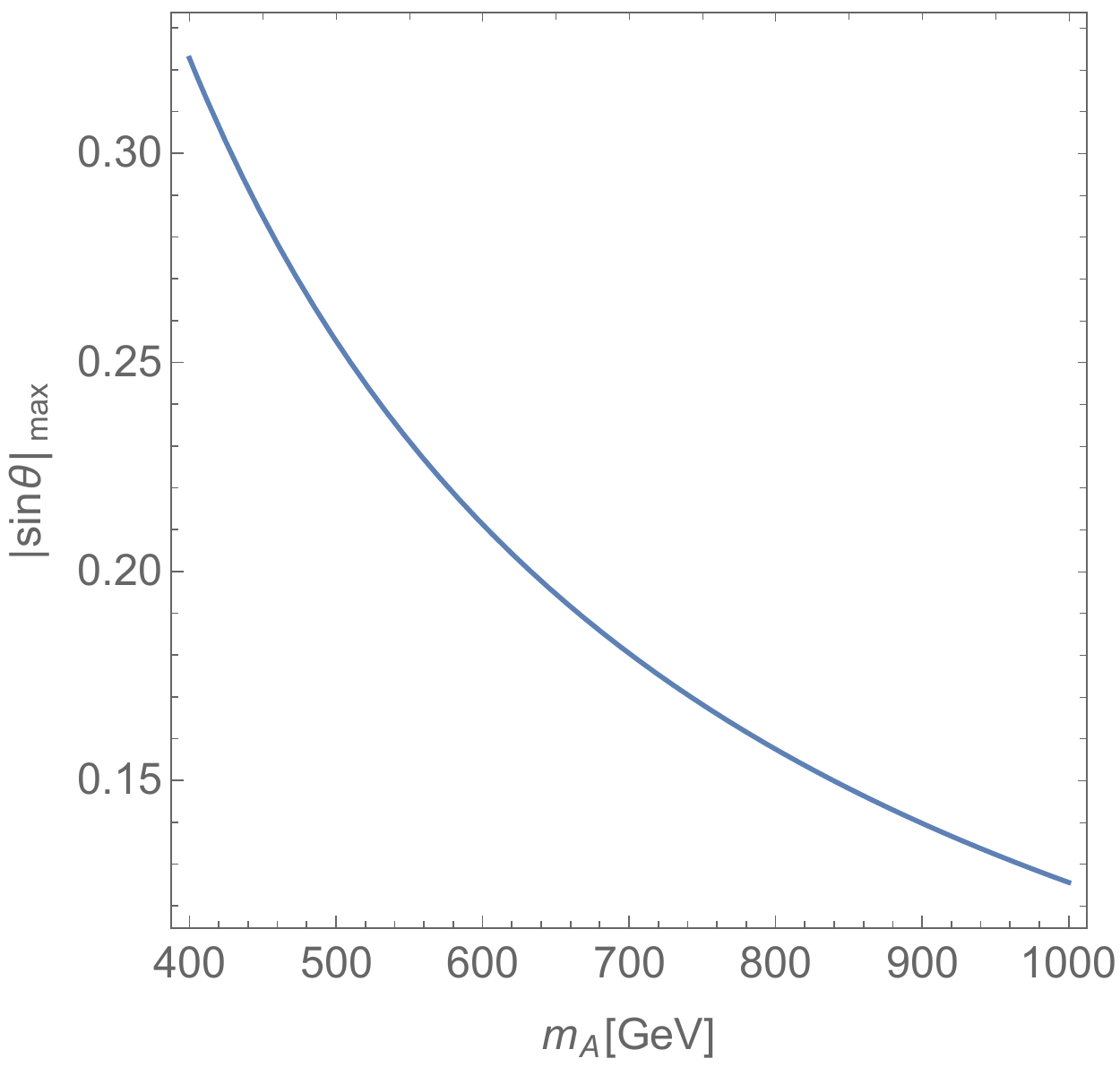}
\caption{
The upper bound on $\theta$ obtained from Eq.~\eqref{eq:max-theta}.
Here we take $s_{\beta-\alpha} = 1$, $m_a = 100$~GeV, and $m_H = m_{H^{\pm}} = m_A$.
}
\label{fig:theta-max}
\end{figure}

We find that Eq.~\eqref{eq:BFB_4} gives a constraint on $\theta$.
For $s_{\beta-\alpha} = 1$ and $M = m_H = m_A = m_{H^\pm}$,
using Eqs.~\eqref{eq:lam123} and \eqref{eq:lam45},
we can simplify Eq.~\eqref{eq:BFB_4} as
\begin{align}
 | \sin\theta | < \frac{m_h}{\sqrt{m_A^2 - m_a^2}}.
\label{eq:max-theta}
\end{align}
The constraint on $|\sin\theta|$ by using this result for $m_a = 100$~GeV is shown in Fig.~\ref{fig:theta-max}.
We find that 
$|\sin\theta| \lesssim 0.21$ for $m_{A} = 600$~GeV, and
$|\sin\theta| \lesssim 0.13$ for $m_{A} = 1$~TeV.

For $s_{\beta-\alpha} = 1$ and $M = m_H = m_A = m_{H^\pm}$, we can also simplify other conditions with physical observables as follows.
\begin{align}
& \sqrt{\frac{\lambda_a}{2}}\frac{m_h}{v} + c_1 > 0, \\
& \sqrt{\frac{\lambda_a}{2}}\frac{m_h}{v} + c_2 > 0, \\
&\begin{cases}
c_1 + c_2\geq 0,\\
\text{or} \\%
c_1 + c_2< 0 
\text{ and } 
\frac{\lambda_a}{2}
\left(
\frac{m_h^2}{v^2} - \frac{2(m_A^2-m_a^2)}{v^2}s_\theta^2
\right)
- c_1 c_2 + \sqrt{\left( \frac{\lambda_a m_h^2}{2v^2} - c_1^2  \right)\left( \frac{\lambda_a m_h^2}{2v^2} - c_2^2  \right)} > 0.
\end{cases}
\end{align}

\subsection{Perturbative unitarity}

Constraints on scalar quartic couplings are often derived from the perturbative unitarity of 
 two scalars to two scalars scattering processes. There are nine scalars in the model. 
Therefore, the two to two scattering matrix that only includes scalars is a $45 \times 45$ matrix.
Since we consider the high energy limit and ignore the gauge couplings, 
the scattering processes are $s$-wave.
In the following analysis, the Yukawa coupling, $g_\chi$ , often takes ${\cal O}(1)$ value,
and thus we also include DM two body initial and final states in the matrix. 
The DM particle takes two helicity states.
In the high energy limit, 
we find that two DM particles into two DM particles processes are $s$-wave,
and two DM particles into two scalars processes are $p$-wave. 
The former processes give stronger bound on $g_\chi$.
We impose absolute values of each eigenvalue of the matrix are less than $8\pi$ 
and find that~\cite{Horejsi:2005da, 1612.01309}
\begin{align}
 |c_1| <& 4\pi, \label{eq:PU-c1}\\
 |c_2| <& 4\pi, \label{eq:PU-c2}\\
 |\lambda_3 \pm \lambda_4| <& 8\pi, \label{eq:PU-3}\\
\left|
\frac{1}{2}
\left( \lambda_1 + \lambda_2 \pm \sqrt{(\lambda_1-\lambda_2)^2 + 4 \lambda_4^2} \right)
\right|
<& 8\pi, \label{eq:PU-4}\\
\left|
\frac{1}{2}
\left( \lambda_1 + \lambda_2 \pm \sqrt{(\lambda_1-\lambda_2)^2 + 4 \lambda_5^2} \right)
\right|
<& 8\pi, \label{eq:PU-5}\\
|\lambda_3 + 2 \lambda_4 \pm 3 \lambda_5| <& 8\pi, \label{eq:PU-6}\\
|\lambda_3 \pm \lambda_5| <& 8\pi, \label{eq:PU-7}\\
g_\chi^2 <& 4\pi,\label{eq:PU-gchi}\\
|x_i| <& 8 \pi \ (i = 1,2,3),
\label{eq:PU-last}
\end{align}
where $x_i$ are solutions of the following equation,
\begin{align}
0=& 
x^3
- 3 \left(\lambda_a + \lambda_1 + \lambda_2 \right) x^2
\nonumber\\
&
+\left(
-4 c_1^2 - 4 c_2^2  - 4 \lambda_3^2 - 
  4 \lambda_3 \lambda_4 - \lambda_4^2 
+ 9 \lambda_1 \lambda_2+ 9 \lambda_1 \lambda_a +   9 \lambda_2 \lambda_a
\right)
x
\nonumber\\
&
+
12 c_2^2 \lambda_1 + 12 c_1^2 \lambda_2 - 16 c_1 c_2 \lambda_3 - 
 8 c_1 c_2 \lambda_4 
+\left(
- 27 \lambda_1 \lambda_2 +  12 \lambda_3^2 + 12 \lambda_3 \lambda_4 + 3 \lambda_4^2 
\right)
\lambda_a
.
\end{align}
For $|\lambda_i| \ll 1 \ (i = 1,2,3,4,5)$, Eq.~\eqref{eq:PU-last} is simplified as
\begin{align}
\frac{1}{2} \left(3 \lambda_a + \sqrt{16 c_1^2 + 16 c_2^2 + 9 \lambda_a^2}\right) <& 8 \pi,
\end{align}
or
\begin{align}
 \lambda_a < \frac{8\pi}{3} \left( 1 - \frac{c_1^2 + c_2^2}{16\pi^2}  \right).
\label{eq:max-lam_a}
\end{align}
Since $\lambda_a >0$, this inequality implies that
\begin{align}
 \sqrt{c_1^2 + c_2^2} < 4 \pi.
\end{align}
This gives stronger constraint on $c_1$ and $c_2$ than Eqs.~\eqref{eq:PU-c1} and \eqref{eq:PU-c2}.

For $s_{\beta-\alpha} = 1$ and $M = m_H = m_A = m_{H^\pm}$,
using Eqs.~\eqref{eq:lam123} and \eqref{eq:lam45}, we can simplify the perturbative unitarity conditions $\lambda_{1,2,3,4,5}$  
(Eqs.~\eqref{eq:PU-3}--\eqref{eq:PU-7})
and express them with masses of the scalars as follows.
\begin{align}
& |m_h^2 \pm (m_A^2 - m_a^2) s_\theta^2| < 8 \pi v^2,\\
& |m_h^2 -5 (m_A^2 - m_a^2) s_\theta^2| < 8 \pi v^2.
\end{align}

\section{Spin-independent scattering cross section}\label{sec:main}

We discuss the maximum value of $\sigma_\text{SI}$ under the constraints discussed in Sec.~\ref{sec:constraints}.
We find upper bounds on $c_1$ and $c_2$ from the stability of the electroweak vacuum,
and lower bounds from the boundedness of the scalar potential.
Since $g_\chi \sim {\cal O}(1)$ in the large $\sigma_\text{SI}$ regime~\cite{1810.01039},
the perturbative unitarity also gives relevant constraint in the parameter space.

\begin{figure}[tb]
\includegraphics[width=0.48\hsize]{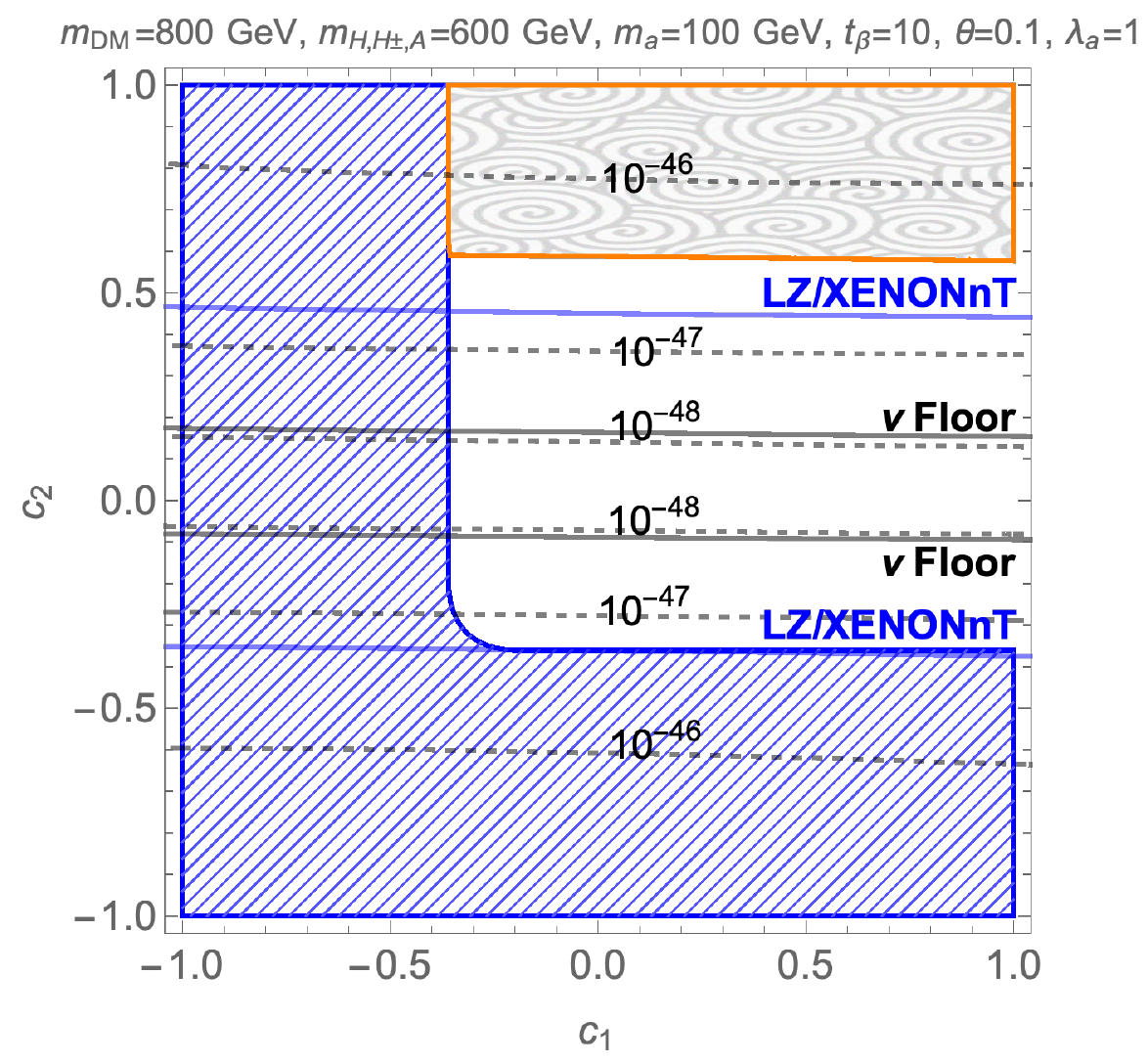} 
\includegraphics[width=0.45\hsize]{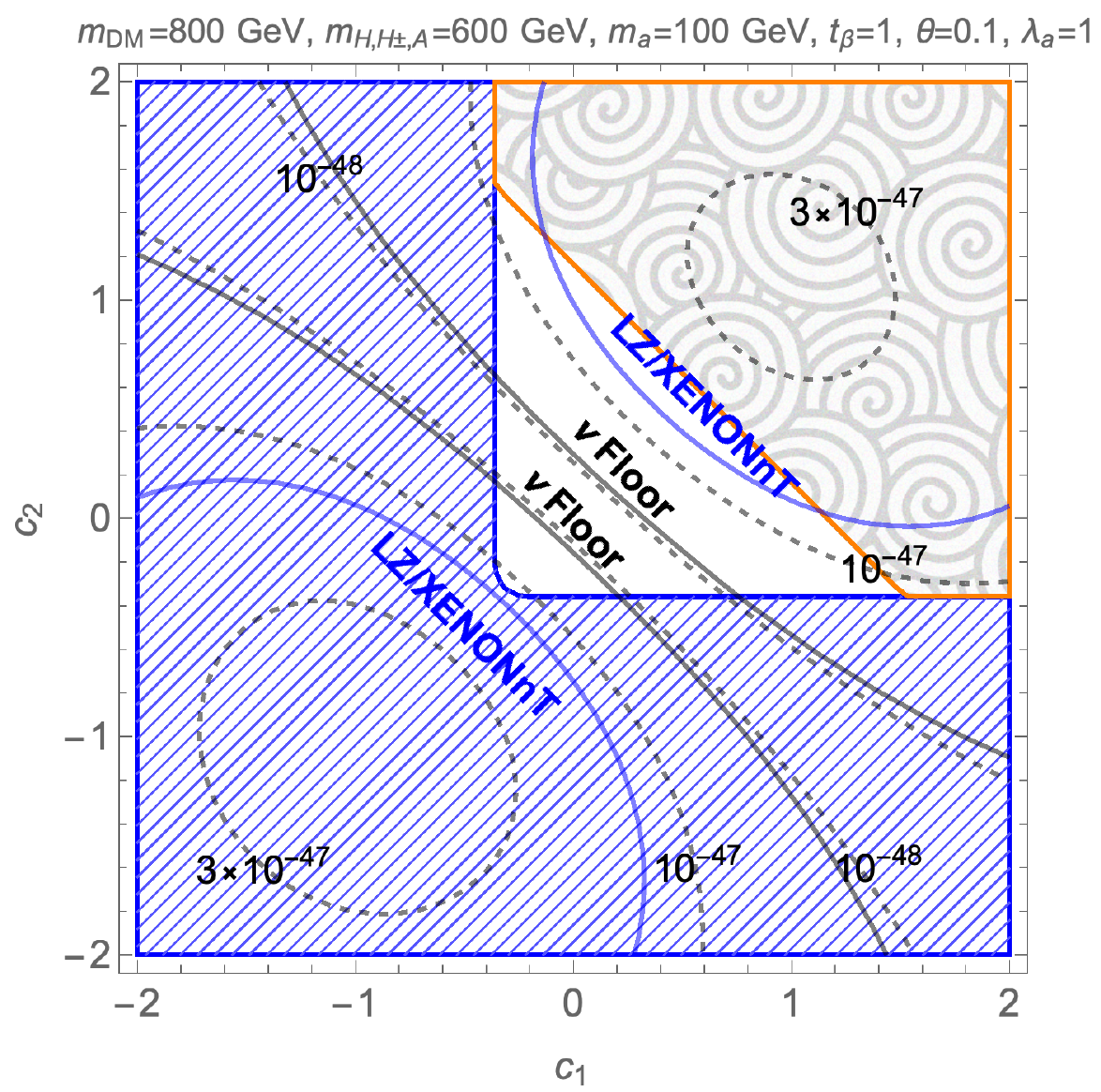}
\caption{
The contours for $\sigma_\text{SI}$ [cm$^2$] for $m_\text{DM} = 800$~GeV are shown by the dashed curves.
The blue solid lines show the LZ and XENONnT prospects~\cite{1802.06039, Aprile:2015uzo},
In both panels, we take $m_{H} = m_{H^\pm} = m_A = 600$~GeV, $m_a = 100$~GeV, $\theta = 0.1$, and $\lambda_a = 1$.
The left (right) panel is for $t_\beta = 10$ (1).
The region between two black solid lines is below the neutrino floor~\cite{1310.8327}.
The global minimum does not break the electroweak symmetry in the region surrounded by the orange line and filled with the spirals pattern.
In the blue shaded region with hatching, the scalar potential is unbounded from the below.
}
\label{fig:mDM800_ma100_mA600_tb10_theta0.1_lama_1.0}
\end{figure}

The left panel in Fig.~\ref{fig:mDM800_ma100_mA600_tb10_theta0.1_lama_1.0} shows that
the contours of $\sigma_\text{SI}$ for $m_\text{DM} = 800$~GeV with
the conditions discussed in Secs.~\ref{sec:global-min} and \ref{sec:bfb}.
The other parameters except $\lambda_a$ are the same as one used in Fig.~8 in Ref.~\cite{1810.01039}, namely $m_{H} = m_{H^\pm} = m_A = 600$~GeV, $m_a = 100$~GeV, $t_\beta =10$, $\theta = 0.1$, and $\lambda_a = 1$.
It is clearly shown that $\sigma_\text{SI}$ is larger in the larger $|c_2|$ region as discussed in Ref.~\cite{1810.01039}.
It is also shown that there is an upper bound on $\sigma_\text{SI}$ from the condition discussed in Sec.~\ref{sec:global-min}.
A large positive $c_2$ predicts that the electroweak vacuum is not the global minimum.
This is because such a large positive $c_2$ makes $m_{a_0}^2$ negatively large as can be seen from Eq.~\eqref{eq:ma02}, and thus Eq.~\eqref{eq:c1c2upperbound} is not satisfied.
A large negative $c_2$ does not satisfy Eqs.~\eqref{eq:BFB_5}--\eqref{eq:BFB_7}
and makes the potential unbounded from the below.
These theoretical constraints on the scalar potential give the upper and lower bounds on $c_2$.
Consequently, $\sigma_\text{SI}$ cannot be arbitrary large.
The right panel in Fig.~\ref{fig:mDM800_ma100_mA600_tb10_theta0.1_lama_1.0} is a similar to the left panel 
but with a smaller value of $t_\beta$. 
In this case, there are upper bounds both on $c_1$ and $c_2$.

From Fig.~\ref{fig:mDM800_ma100_mA600_tb10_theta0.1_lama_1.0},
we find a correlation between $\sigma_\text{SI}$ and the condition of the stability of the electroweak vacuum. 
The contour of $\sigma_\text{SI}$ and the boundary of the constraint of the stability of the electroweak vacuum (the edge of the orange shaded region) are almost parallel to each other.
We can understand this correlation as follows.
For $s_{\beta-\alpha} =1$ and $m_{H}=m_{H^{\pm}}$, 
the condition to avoid $\VEV{a_0} \neq 0$ vacuum given in Eq.~\eqref{eq:No-a0-vev} is simplified as
\begin{align}
 \lambda_a >  \frac{2 m_{a_0}^4}{m_h^2 v^2}
\quad \text{   (for $m_{a_0}^2 < 0$)}
. 
\label{eq:No-a0-vev-no2}
\end{align}
As discussed in Ref.~\cite{1810.01039}, both $\sigma_\text{SI}$ and $\VEV{\sigma v}$ depend on the $a$-$a$-$h$ coupling, $g_{aah}$, that is given by
\begin{align}
 g_{aah}
=&
s_\theta^2 \left(
  \frac{2 m_a^2 + m_h^2 - 2 m_A^2}{v}
\right)
+ 
2 v c_\theta^2 
  \frac{c_1 + c_2 t_\beta^2}{1+t_\beta^2}
\label{eq:g_aah}
\\
=&
\frac{2 (m_a^2 - m_{a_0}^2)}{v}
+ {\cal O}(\theta^2)
,
\end{align} 
for $s_{\beta-\alpha} =1$ and $m_{H}=m_{H^{\pm}} = m_A$. 
Combining these two equations, we find
\begin{align}
\lambda_a
>&
 \frac{2 v^2}{ m_h^2}
\left(
  \frac{m_a^2}{v^2}
 -\frac{g_{aah}}{2v} 
\right)^2
+
{\cal O}\left( \theta^2 \right)
.
\label{eq:lama_vs_gaah}
\end{align}
This condition is not satisfied with the large $g_{aah}$,
and thus the large $g_{aah}$ induces the $\VEV{a_0} \neq 0$ vacuum.
On the other hand, the large $g_{aah}$ is necessary to obtain the larger $\sigma_\text{SI}$.
Therefore,
there is a correlation between $\sigma_\text{SI}$ and the condition of the stability of the electroweak vacuum.

We can also see from Fig.~\ref{fig:mDM800_ma100_mA600_tb10_theta0.1_lama_1.0} that
the maximum value of $\sigma_\text{SI}$ is near the boundary of the stability of the electroweak vacuum. 
For the purpose of finding maximum value of $\sigma_\text{SI}$, 
we need to find the maximum value of $g_{aah}$ that satisfies Eq.~\eqref{eq:lama_vs_gaah}.
The $c_1$ and $c_2$ dependent part of $g_{aah}$, which is the second term in Eq.~\eqref{eq:g_aah}, 
depends on $t_\beta$.
This $t_\beta$ dependence vanishes for $c_1 = c_2$.
We take $c_1 = c_2$ and $t_\beta = 10$ in the following analysis, 
but the following results are insensitive to the choice of $t_\beta$.

The larger $\lambda_a$ allows us to take larger $g_{aah}$ while keeping $\VEV{a_0} = 0$,
which can be seen from Eq.~\eqref{eq:lama_vs_gaah}.
On the other hand, 
the larger $\lambda_a$ implies the breakdown of perturbative calculation at a higher energy scale.
In our analysis, $g_\chi$ is typically ${\cal O}(1)$ to obtain the measured value of the DM energy density, and it also implies the breakdown of perturbative calculation at a higher energy scale.
We calculate the running of the couplings at the 1-loop level 
and estimate the cutoff scale $\Lambda$ as the highest scale that satisfies
Eqs.~\eqref{eq:BFB_3}, \eqref{eq:PU-gchi}, and \eqref{eq:max-lam_a}.
In the calculation, we assume that the input parameters are given at $\mu = m_A$. 
The beta-functions of the couplings we used are given in Appendix~\ref{sec:beta-function}.
The smaller $\lambda_a$ at $\mu = m_A$ becomes negative at higher scale because $a_0$ couple to the fermionic DM that gives a negative contribution to the beta function of $\lambda_a$.
On the other hand, the beta function is proportional to $\lambda_a^2$ and positive for the larger $\lambda_a$.
The cutoff scale gives the upper and the lower bounds on $\lambda_a$ at $\mu = m_A$.

\begin{figure}[tb]
\includegraphics[width=0.48\hsize]{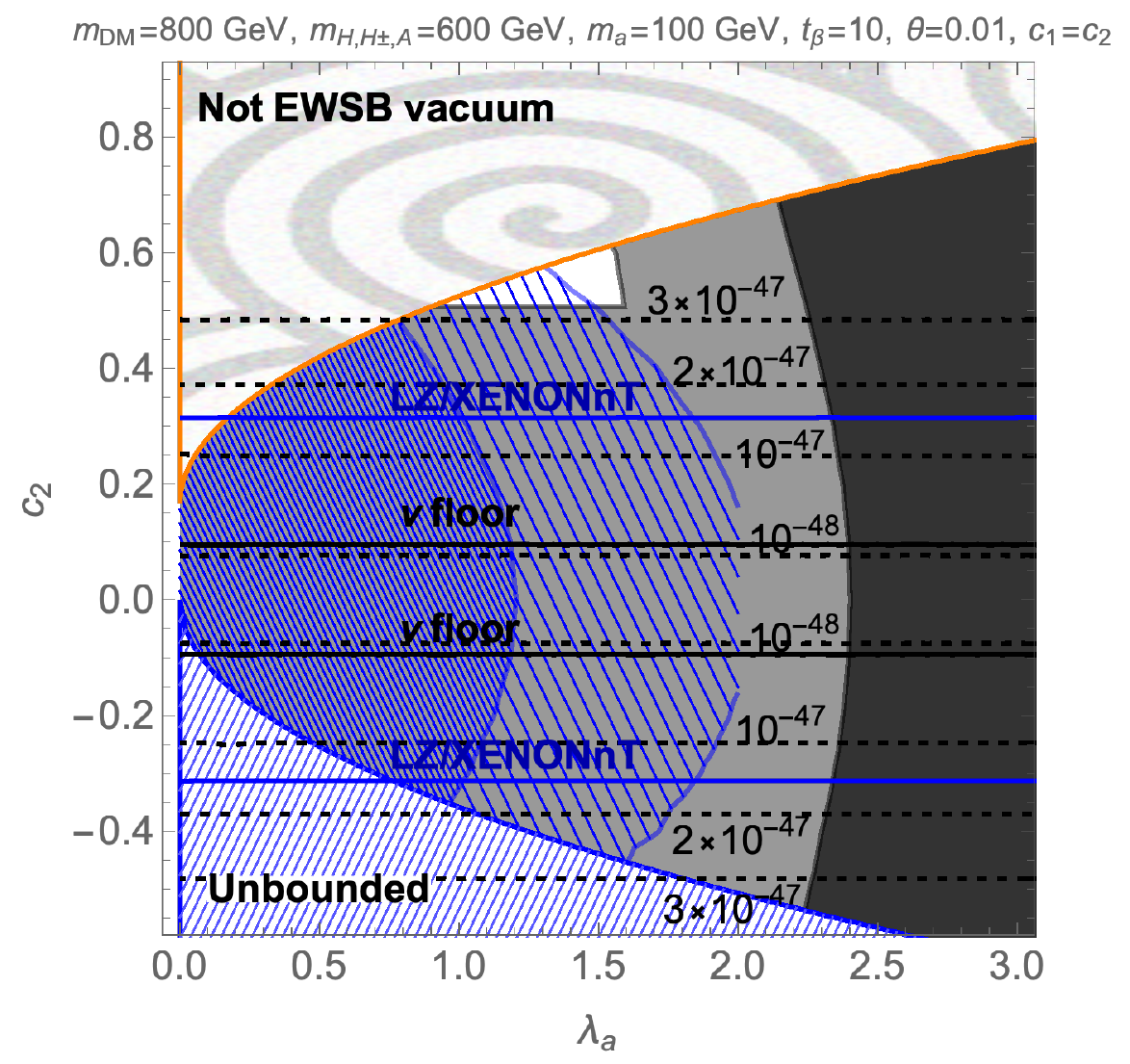}
\includegraphics[width=0.48\hsize]{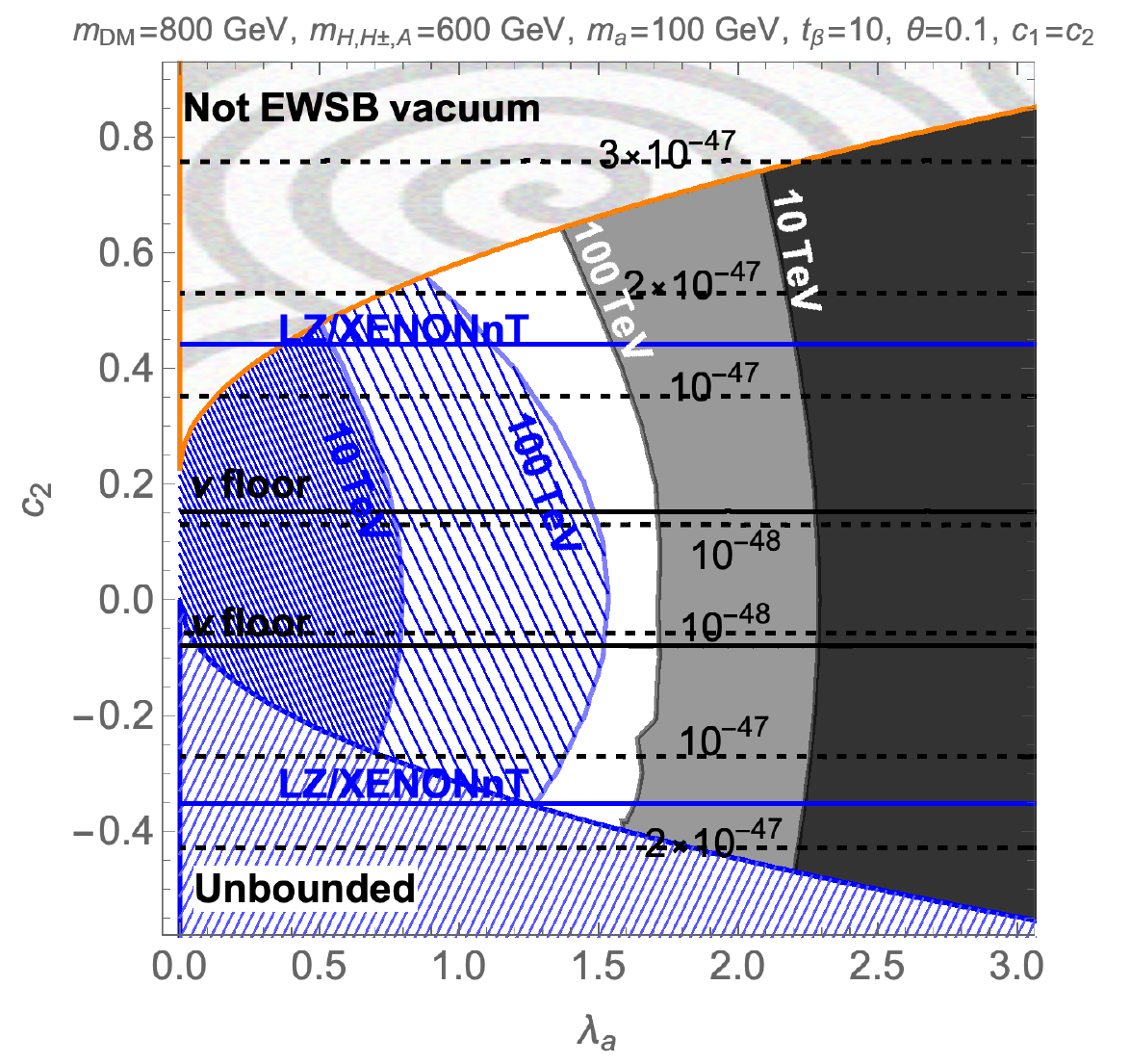}
\caption{
The contours for $\sigma_\text{SI}$ [cm$^2$] for $m_\text{DM} = 800$~GeV are shown by the dashed curves.
In the left (right) panel, $\theta = 0.01$ (0.1).
In the thin (dense) blue-hatching region (\textbackslash \textbackslash \textbackslash), $\lambda_a(\Lambda)$ becomes negative at $\Lambda < 100~(10)$~TeV.
In the lighter (darker) black region, Eq.~\eqref{eq:PU-gchi} or \eqref{eq:max-lam_a} is violated at $\Lambda < 100~(10)$~TeV.
The other color notation is the same as in Fig.~\ref{fig:mDM800_ma100_mA600_tb10_theta0.1_lama_1.0}.
}
\label{fig:lam_a-vs-c2}
\end{figure}

Figure~\ref{fig:lam_a-vs-c2} shows the contours of $\sigma_\text{SI}$ in $\lambda_a$-$c_2$ planes.
It is shown that the larger $\lambda_a$ at $\mu = m_A$ keeps its value positive at any higher scale.
We find that 
it is easy to make the cutoff scale higher than ${\cal O}(100)$~TeV by choosing $\lambda_a \simeq 1.5$.
Thus we can expect that unknown UV physics does not modify our results for $\lambda_a \simeq 1.5$.
We also find that
$\sigma_\text{SI}$ is maximized along the boundary of the orange shaded region where the electroweak symmetry is not broken.
For $c_1 = c_2$, Eq.~\eqref{eq:c1c2upperbound} is simplified as
\begin{align}
c_2<
 \sqrt{\frac{\lambda_a m_h^2}{2 v^2}} 
+\frac{m_a^2 c_\theta^2 + m_A^2 s_\theta^2}{v^2} 
\equiv c_*
.
\label{eq:cstar}
\end{align}
In the following analysis, we choose $c_1 = c_2 = 0.99 c_*$ for given parameter sets.
This choice of $c_1$ and $c_2$ maximizes $\sigma_\text{SI}$.

\begin{figure}[tb]
\includegraphics[width=0.48\hsize]{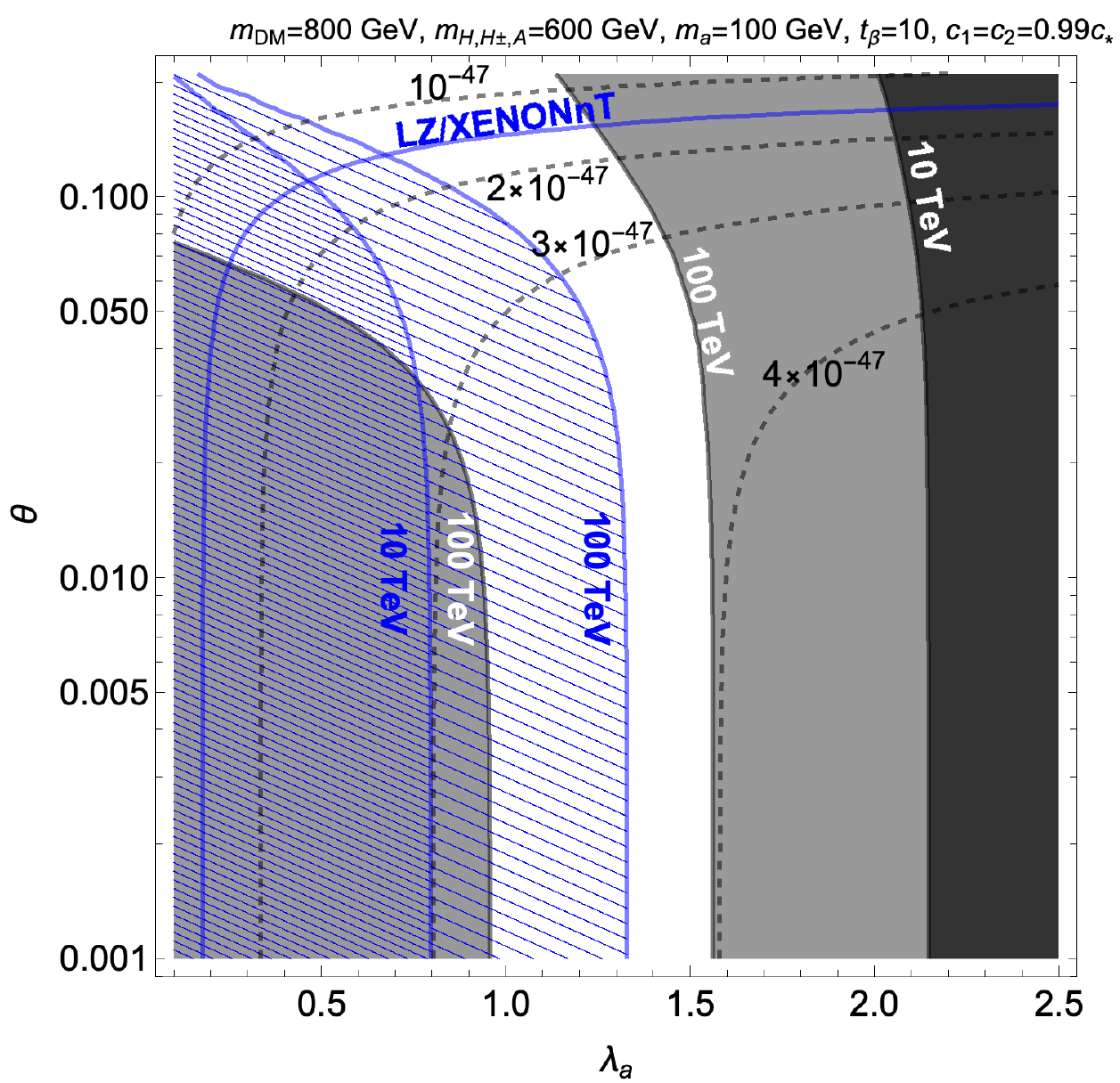}
\includegraphics[width=0.48\hsize]{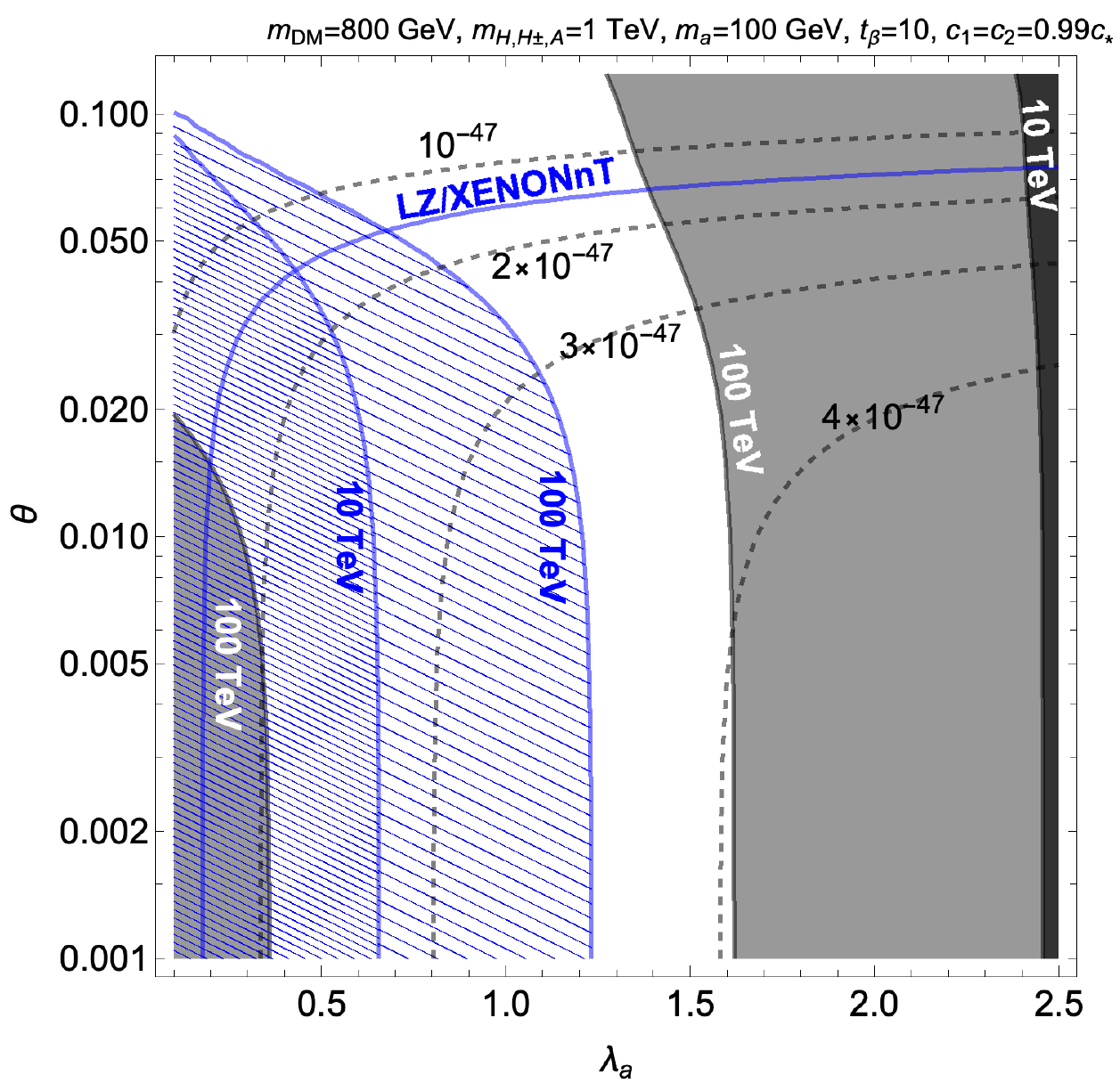}
\caption{
The contours for $\sigma_\text{SI}$ [cm$^2$] in $\lambda_a$-$\theta$ plane.
The left (right) panel is for $m_A = $600 (1000)~GeV.
The color notation is the same in Fig.~\ref{fig:lam_a-vs-c2}.
}
\label{fig:lam_a-vs-thata}
\end{figure}

Figure~\ref{fig:lam_a-vs-thata} shows the contours of $\sigma_\text{SI}$ in $\lambda_a$-$\theta$ plane.
We find that $\sigma_\text{SI}$ is larger in the smaller $\theta$ regime.
This is because the smaller $\theta$ requires larger $g_\chi$ to obtain the measured
value of the DM energy density.
The left and right panels are for $m_A = 600$~GeV and 1~TeV, respectively.
We find that $\sigma_\text{SI}$ is almost independent from $m_{H, H^\pm, A}$ for $\theta < 0.01$. This is because the heavier scalars almost decouple both from the DM annihilation 
processes and from the loop contributions to $\sigma_\text{SI}$.
The cutoff scales are the only difference if we change $m_A$;
a larger $m_A$ predicts higher cutoff scales.
In the following analysis, we take $\theta = 0.001$.
With this choice, $\sigma_\text{SI}$ is maximized and is independent from $m_A$.
We also take $m_A = 600$~GeV in the following, which gives us a conservative bound from the RGE analysis.

\begin{figure}[tb]
\includegraphics[width=0.48\hsize]{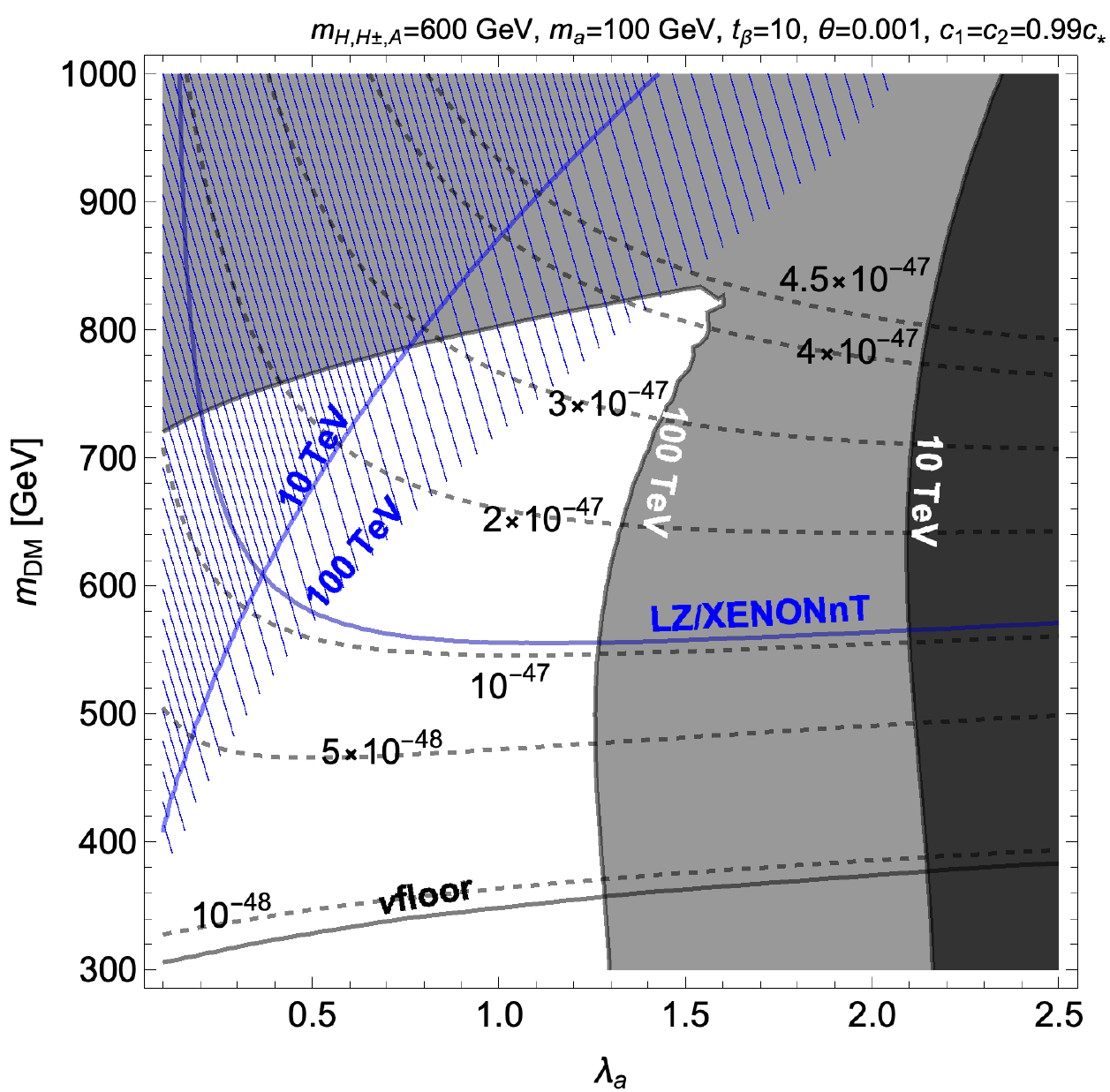}
\includegraphics[width=0.48\hsize]{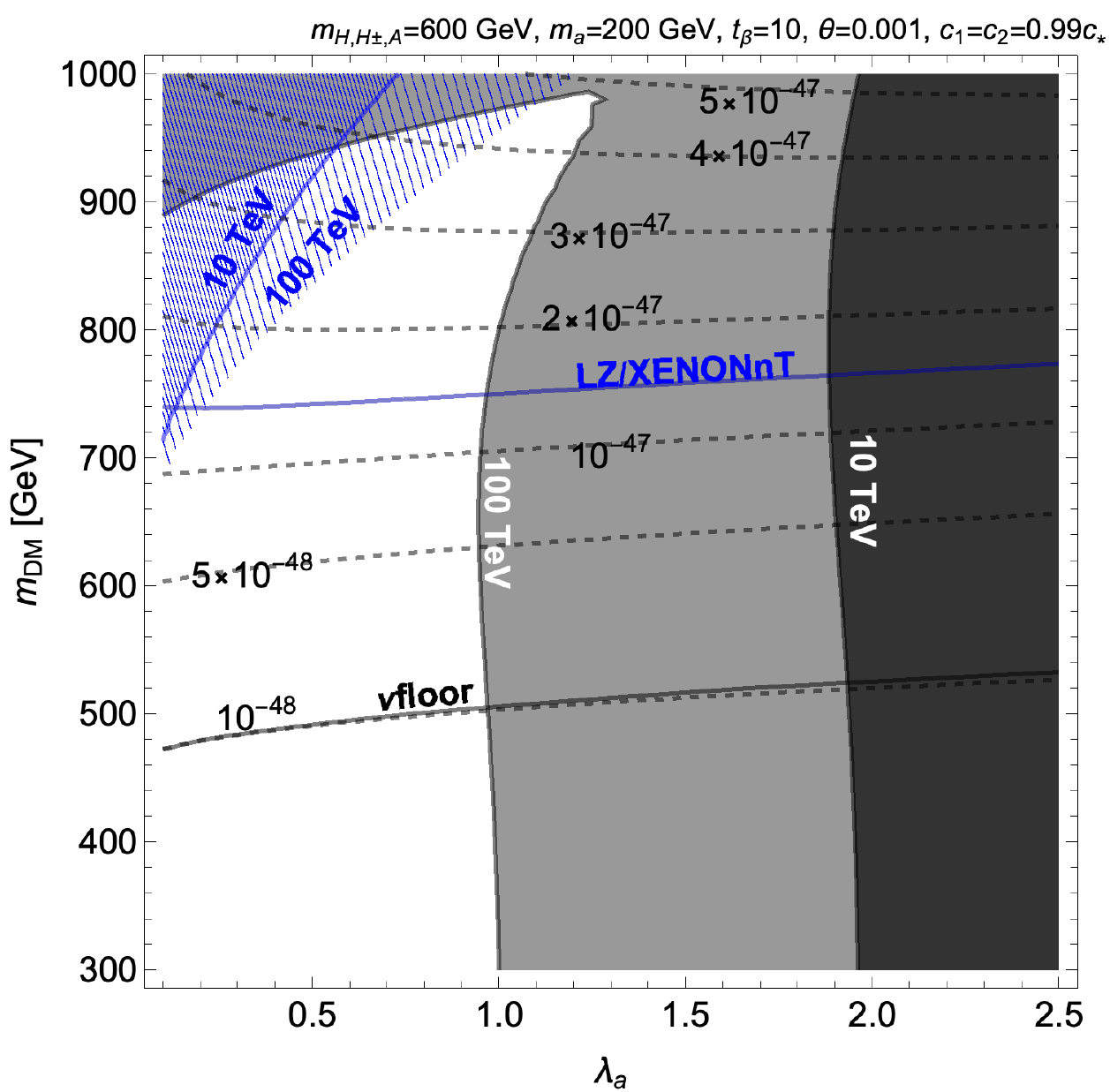}
\includegraphics[width=0.48\hsize]{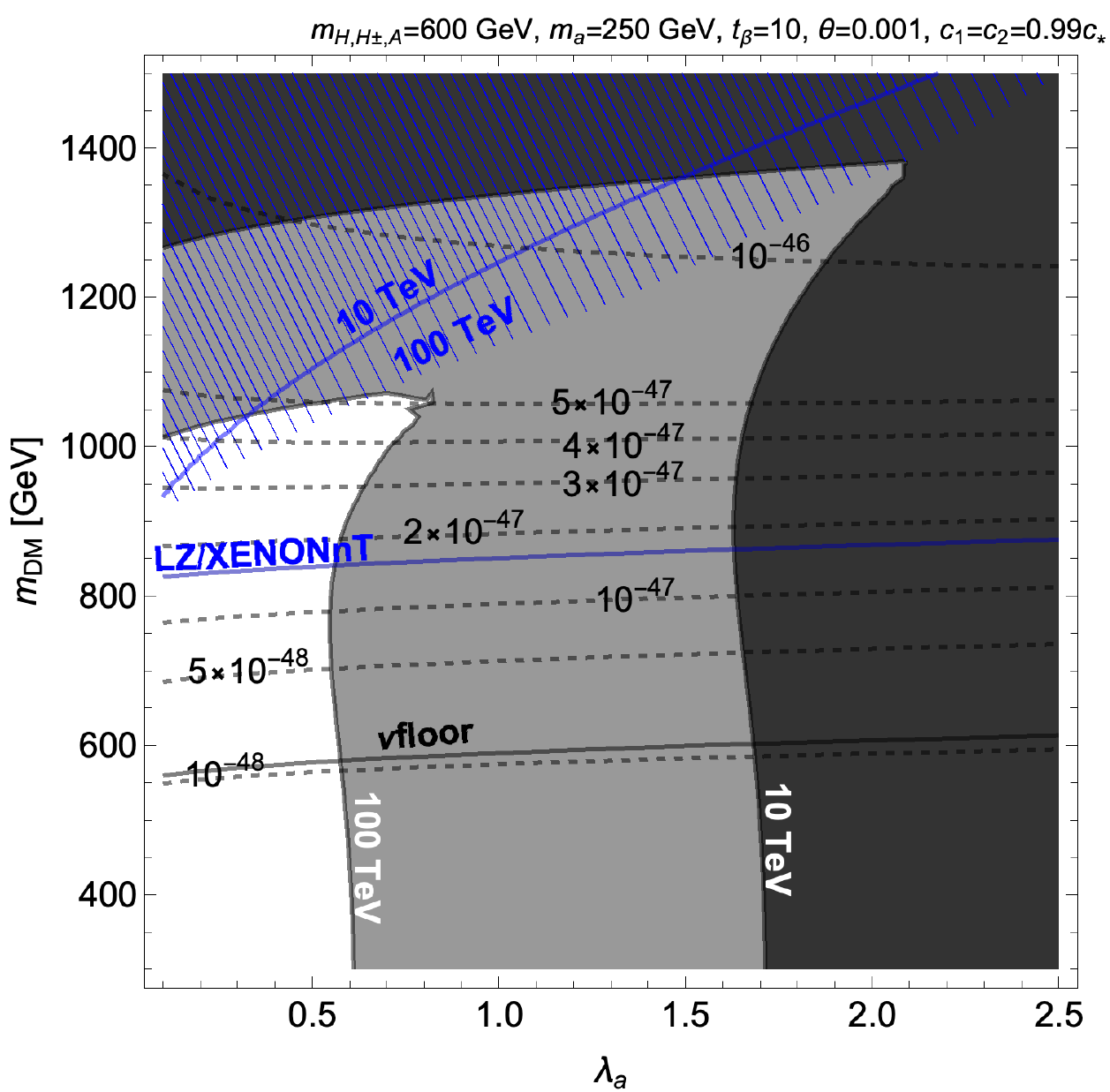}
\includegraphics[width=0.48\hsize]{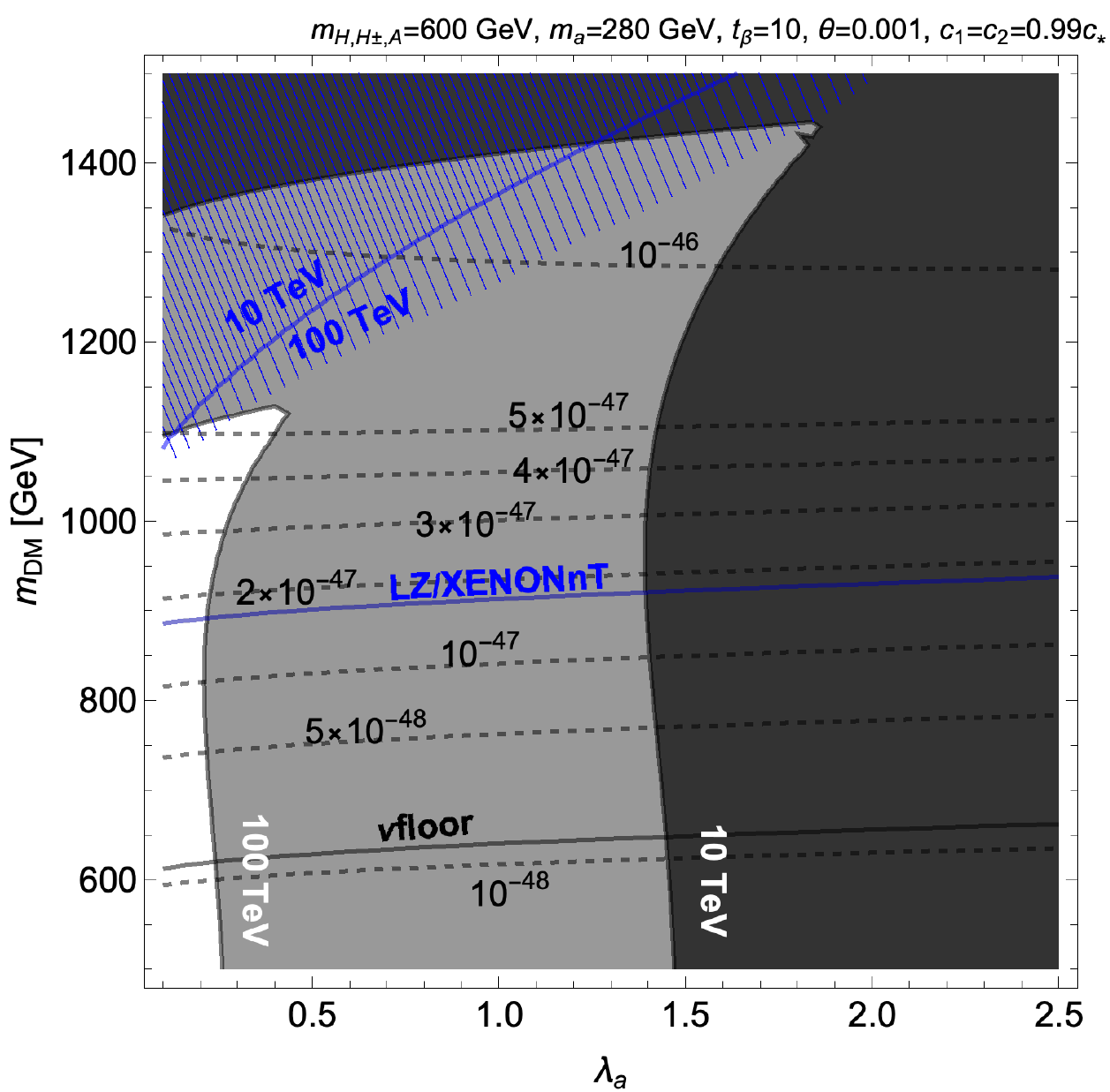}
\caption{
The contours for $\sigma_\text{SI}$ [cm$^2$] in $\lambda_a$-$m_\text{DM}$ plane.
We take $m_a = 100$, 200, 250, and 280~GeV in each panel.
The color notation is the same in Fig.~\ref{fig:lam_a-vs-c2}.
}
\label{fig:lam_a-vs-mDM}
\end{figure}

Figure~\ref{fig:lam_a-vs-mDM} shows the contours of $\sigma_\text{SI}$ in $\lambda_a$-$m_\text{DM}$ plane.
We take $m_a = 100$, 200, 250, and 280~GeV in each panel.
We find that the maximum value of $\sigma_\text{SI}$ is almost independent of the choice of $m_a$,
$\sigma_\text{SI} \lesssim 5 \times 10^{-47}$~cm$^2$. 
This value is larger than the prospects of the LZ and XENONnT experiments. 
Therefore, we have a chance to see the DM direct detection signal near future.
The constraint from the perturbative unitarity with the running couplings gives a stronger bound for the larger $m_a$ due to the following reason. 
As can be seen from Eq.~\eqref{eq:cstar}, $c_*$ and hence $c_1$ and $c_2$ 
become larger for the larger $m_a$.
The larger $c_1$ and $c_2$ make the beta function of $\lambda_a$ larger.
Therefore, the constraint from the perturbative unitarity with the running couplings becomes severer for larger values of $m_a$.
It is also shown that $\sigma_\text{SI}$ becomes large for the large $m_\text{DM}$ regime.
This is because larger values of $m_\text{DM}$ requires larger values of $g_\chi$ to obtain the right amount of
the relic abundance. On the other hand, larger values of $g_\chi$ implies that the Landau pole arises at a lower scale because $g_\chi$ is asymptotic non-free. This gives an upper bound on $m_\text{DM}$ as shown in the figure.

\section{Conclusion}\label{sec:summary}
The 2HDM+a is a DM model that can explain the measured value of the DM energy density by the freeze-out mechanism and can avoid the constraint from the Xenon1T experiment.
The leading order contribution to $\sigma_\text{SI}$ is given at the loop level, 
and $\sigma_\text{SI}$ can be large enough for the model to be tested by the forthcoming direct detection experiments.

In this paper, we have investigated the maximum value of $\sigma_\text{SI}$ under theoretical constraints. We take into account the stability of the electroweak vacuum,
the condition for the potential bounded from the below,
and the perturbative unitarity of two to two scattering processes.
As shown in Fig.~\ref{fig:mDM800_ma100_mA600_tb10_theta0.1_lama_1.0},
large $|c_1|$ and $|c_2|$ make $\sigma_\text{SI}$ larger.
However, the condition for the stability of the electroweak vacuum gives upper bounds on $c_1$ and $c_2$, and the potential boundedness condition gives lower bounds on them.
As a result, there exists the maximum value of $\sigma_\text{SI}$ for a given parameter set.
It is also shown that $\sigma_\text{SI}$ is maximized for $c_1 = c_2 = c_*$, where $c_*$ is 
the maximum values of $c_1$ and $c_2$ to keep the electroweak vacuum as the global minimum of the scalar potential. With this choice, the result is insensitive to $t_\beta$.
We found that 
a smaller $\theta$ makes $\sigma_\text{SI}$ larger, as shown in Fig.~\ref{fig:lam_a-vs-thata}.
For the small $\theta$ regime, $\sigma_\text{SI}$ is almost independent of $m_{H, H^{\pm},A}$.
Finally, in Fig.~\ref{fig:lam_a-vs-mDM},
we found that $\sigma_\text{SI}$ can be larger than the prospects of the LZ and XENONnT experiments for
$m_\text{DM} \gtrsim 600$~GeV. 
We also found that the perturbative unitarity gives an upper bound on $m_\chi$.
The maximum value of the $\sigma_\text{SI}$ is $\sim 5\times 10^{-47}$~cm$^2$ for
$m_A = $~600~GeV where the cutoff scale of this model is estimated as 100~TeV.
Therefore, if the LZ and XENONnT experiments observe the DM signal in future,
then this model predicts 600~GeV $\lesssim m_\text{DM} \lesssim$~1TeV.

\section*{Acknowledgments}
This work was supported by JSPS KAKENHI Grant Number 16K17715, 18H04615~[T.A.]
and by Grant-in-Aid for Scientific research from the Ministry of
Education, Science, Sports, and Culture (MEXT), Japan, No. 16H06492~[J.H. and Y.S.]. 
The work of J.H. is also supported by World Premier
International Research Center Initiative (WPI Initiative), MEXT, Japan.

\clearpage
\appendix
\renewcommand{\thesubsection}{\Alph{section}.\arabic{subsection}}
\section{Condition for the potential to be bounded below}
\label{app:BFB}
The potential should be bounded below, 
namely the potential should be positive for the region where the field values are extremely larger.
In this section, we derive the condition for the bounded below.

We focus on the region where the fields take large values,
and thus the quadratic and cubic terms in the potential are negligible in 
the analysis here,
\begin{align}
 V\sim&
+ \frac{1}{2} \lambda_1 (H_1^{\dagger} H_1)^2
+ \frac{1}{2} \lambda_2 (H_2^{\dagger} H_2)^2
+ \lambda_3 (H_1^{\dagger} H_1) (H_2^{\dagger} H_2)
+ \lambda_4 (H_1^{\dagger} H_2) (H_2^{\dagger} H_1)
\nonumber\\
&
+
\frac{1}{2}
\lambda_5
\left(
 (H_1^{\dagger} H_2)^2 
+
(h.c.)
\right)
+\frac{\lambda_a}{4} a_0^4
+ c_1 a_0^2 H_1^\dagger H_1
+ c_2 a_0^2 H_2^\dagger H_2
.
\end{align}
We introduce the following parametrization,
\begin{align}
H_1^\dagger H_1 =& \rho^2 \sin\theta \cos\phi,\\
H_2^\dagger H_2 =& \rho^2 \sin\theta \sin\phi,\\
H_1^\dagger H_2 =& \rho^2 \sin\theta \sqrt{\cos\phi \sin\phi } e^{-i \theta_3} \cos\omega,\\
a_0^2 =& \rho^2 \cos\theta,
\end{align}
where $\rho^2 >0$, $0 \leq \theta \leq \pi/2$, and $0 \leq \phi \leq \pi/2$.
Using these parameters, the scalar potential is written as
\begin{align}
 \frac{V}{\rho^4}
\sim 
 \tilde{V}
\equiv&
  \frac{1}{2} \lambda_1 \sin^2\theta \cos^2\phi
+ \frac{1}{2} \lambda_2 \sin^2\theta \sin^2\phi
\nonumber\\
&
+ \frac{1}{2}
\left[ \lambda_3 + \left( \lambda_4 + \lambda_5 \cos\theta_3 \right) \cos^2\omega  \right]
\sin^2\theta \sin(2\phi)
\nonumber\\
&
+ \frac{\lambda_a}{4} \cos^2\theta
+ \frac{1}{2} c_1 \sin(2\theta) \cos\phi
+ \frac{1}{2} c_2 \sin(2\theta) \sin\phi
.
\label{eq:Vtilde}
\end{align}
By imposing $\tilde{V} > 0$, we find constraints on the parameters.

There is a relation we will use in the rest of this section.
Assume $a >0$, $b>0$, and $0 < \theta < \pi/2$, then
\begin{align}
  a \cos^2\theta + b\sin^2\theta + 2 c \sin\theta \cos\theta > 0
\end{align}
if $c + \sqrt{ab} >0$.
The proof is the following.
\begin{align}
 a \cos^2\theta + b\sin^2\theta + 2 c \sin\theta \cos\theta  
=&
 \left( \sqrt{a} \cos\theta \pm \sqrt{b} \sin\theta \right)^2
+ 2 \sin\theta \cos\theta
\left( c \mp \sqrt{ab}\right)
\nonumber\\
=&
2 \sin\theta \cos\theta
\left(
 \frac{\left( \sqrt{a} \cos\theta \pm \sqrt{b} \sin\theta \right)^2}{2 \sin\theta \cos\theta}
+ 
\left( c \mp \sqrt{ab}\right)
\right)
.
\end{align}
The sign of the left-hand side is determined by the sign of the terms in the big parenthesis in
the right-hand side. It takes minimum if the terms depending on $\theta$ vanish,
namely, $\sqrt{a} \cos\theta - \sqrt{b} \sin\theta =0$. Its minimum value is
$c + \sqrt{ab}$. Since $\sin\theta \cos\theta > 0$, if $c+ \sqrt{ab} >0$ then
the right-hand side is always positive.

\subsection{ $\theta =0$}
For $\theta =0$, which is the case for $H_1 = H_2 = 0$, we find
\begin{align}
 \tilde{V}
=&
+ \frac{\lambda_a}{4} 
.
\end{align}
Therefore, $\lambda_a >0$.
This is Eq.~\eqref{eq:BFB_3}.

\subsection{$\theta = \pi/2$}
For $\theta = \pi/2$, the potential is the same as in the 2HDMs.
\begin{align}
 \tilde{V}
=&
  \frac{1}{2} \lambda_1 \cos^2\phi
+ \frac{1}{2} \lambda_2  \sin^2\phi
+ \frac{1}{2}
\left[ \lambda_3 + \left( \lambda_4 + \lambda_5 \cos\theta_3 \right) \cos^2\omega  \right]
\sin(2\phi)
.
\end{align}
This potential is simplified for $\phi = 0$ and $\pi/2$,
\begin{align}
 \tilde{V}
=&
\begin{cases}
  \frac{1}{2} \lambda_1  & \text{ ($\phi = 0$)} \\
  \frac{1}{2} \lambda_2  & \text{ ($\phi = \pi/2$)} \\
\end{cases}
.
\end{align}
Therefore, $\lambda_1 >0$ and $\lambda_2 >0$ are required.
These are Eqs.~\eqref{eq:BFB_1} and ~\eqref{eq:BFB_2}.

For $\theta = \pi/2$ and $0 < \phi < \pi/2$,
the potential is positive if
\begin{align}
 \sqrt{\lambda_1 \lambda_2} 
+
\left[ \lambda_3 + \left( \lambda_4 + \lambda_5 \cos\theta_3 \right) \cos^2\omega  \right]
 >0.
\end{align}
We can simplify this inequality.
If 
\begin{align}
\lambda_4 + \lambda_5 \cos\theta_3 \geq 0,
\end{align}
 then  
\begin{align}
\lambda_3 + \left( \lambda_4 + \lambda_5 \cos\theta_3 \right) \cos^2\omega
\geq 
\lambda_3, 
\end{align}
and thus
\begin{align}
  \sqrt{\lambda_1 \lambda_2} 
+
\left[ \lambda_3 + \left( \lambda_4 + \lambda_5 \cos\theta_3 \right) \cos^2\omega  \right]
 \geq 
  \sqrt{\lambda_1 \lambda_2}  + \lambda_3
.
\end{align}
If 
\begin{align}
\lambda_4 + \lambda_5 \cos\theta_3 < 0,
\end{align}
 then  
\begin{align}
\lambda_3 + \left( \lambda_4 + \lambda_5 \cos\theta_3 \right) \cos^2\omega
\geq 
\lambda_3 + \left( \lambda_4 + \lambda_5 \cos\theta_3 \right) 
\geq 
\lambda_3 + \left( \lambda_4 - |\lambda_5| \right) 
,
\end{align}
and thus
\begin{align}
  \sqrt{\lambda_1 \lambda_2} 
+
\left[ \lambda_3 + \left( \lambda_4 + \lambda_5 \cos\theta_3 \right) \cos^2\omega  \right]
 \geq 
  \sqrt{\lambda_1 \lambda_2} 
+
\left[ \lambda_3 + \left( \lambda_4 - |\lambda_5| \right) \right]
.
\end{align}
As a result, we can simplify 
$  \sqrt{\lambda_1 \lambda_2} 
+
\left[ \lambda_3 + \left( \lambda_4 + \lambda_5 \cos\theta_3 \right) \cos^2\omega  \right]
 >0$
as 
\begin{align}
\sqrt{\lambda_1 \lambda_2} 
+
\lambda_3 + \text{min}\left( 0, \lambda_4 - |\lambda_5| \right) 
>0
.
\end{align}
This is Eqs.~\eqref{eq:BFB_4} and 
the same as a condition given in the 2HDMs with softly broken $Z_2$ symmetry~\cite{
PhysRev.D18.2574,
PhysRept.179.273,
hep-ph/9811234,
hep-ph/9903289}.

\subsection{$\phi=0$ and $ 0 < \theta < \pi/2$}
For $\phi=0$ and $ 0 < \theta < \pi/2$,
which is the direction along $H_2 = 0$,
we find 
\begin{align}
 \tilde{V}
=&
  \frac{1}{2} \lambda_1 \sin^2\theta
+ \frac{\lambda_a}{4} \cos^2\theta
+ \frac{1}{2} c_1 \sin(2\theta) 
.
\end{align}
Since $\lambda_1>0$ and $\lambda_a>0$ are already guaranteed, this is positive if 
\begin{align}
 c_1 + \sqrt{\frac{\lambda_1 \lambda_a}{2}} > 0.
\end{align}
This is Eqs.~\eqref{eq:BFB_5}.

\subsection{$\phi=\pi/2$ and $ 0 < \theta < \pi/2$}
For $\phi=\pi/2$ and $0 < \theta<\pi/2$,
which is the direction along $H_1 = 0$,
we find
\begin{align}
 \tilde{V}
=&
+ \frac{1}{2} \lambda_2 \sin^2\theta 
+ \frac{\lambda_a}{4} \cos^2\theta
+ \frac{1}{2} c_2 \sin(2\theta) 
.
\end{align}
Since $\lambda_2>0$ and $\lambda_a>0$ are already guaranteed, 
this is positive if 
\begin{align}
 c_2 + \sqrt{\frac{\lambda_2 \lambda_a}{2}} > 0.
\label{eq:c2-lama}
\end{align}
This is Eqs.~\eqref{eq:BFB_6}.

\subsection{$0 < \phi < \pi/2$ and $ 0 < \theta  < \pi/2$}
For $0 < \phi < \pi/2$ and $ 0 < \theta  < \pi/2$, we need some algebra.
First of all, we can rewrite $\tilde{V}$ as
\begin{align}
 \tilde{V}
=&
\Biggl(
  \frac{1}{2} \lambda_1  \cos^2\phi
+ \frac{1}{2} \lambda_2  \sin^2\phi
+ \frac{1}{2}
\left[ \lambda_3 + \left( \lambda_4 + \lambda_5 \cos\theta_3 \right) \cos^2\omega  \right]
 \sin(2\phi)
\Biggr) \sin^2\theta
\nonumber\\
&
+ \frac{\lambda_a}{4} \cos^2\theta
+ \left( c_1  \cos\phi  + c_2 \sin\phi \right) \sin\theta \cos\theta
.
\end{align}
Since we have already discussed the positivity of $\tilde{V}$ for $\theta = 0$ and $\pi/2$,
we can assume the coefficients of $\cos^2\theta$ and $\sin^2\theta$ are positive. Then,
$\tilde{V}$ is positive if
\begin{align}
\left( c_1  \cos\phi  + c_2 \sin\phi \right) 
+
\sqrt{
 \Biggl(
\lambda_1  \cos^2\phi
+ \lambda_2  \sin^2\phi
+
\left[ \lambda_3 + \left( \lambda_4 + \lambda_5 \cos\theta_3 \right) \cos^2\omega  \right]
 \sin(2\phi)
\Biggr) 
\frac{\lambda_a}{2}
}
>0
.
\end{align}
This should be true for all $\theta_3$ and $\omega$. There for, the following inequality should be satisfied,
\begin{align}
\left( c_1  \cos\phi  + c_2 \sin\phi \right) 
+
\sqrt{
 \Biggl(
\lambda_1  \cos^2\phi
+ \lambda_2  \sin^2\phi
+
\tilde{\lambda}_3
 \sin(2\phi)
\Biggr) 
\frac{\lambda_a}{2}
}
>0
,
\label{sec:bb-a2HDM-3}
\end{align}
where
\begin{align}
\tilde{\lambda}_3 = \lambda_3+ \min(0, \lambda_4 - |\lambda_5|).
\end{align}
Eq.~\eqref{sec:bb-a2HDM-3} is satisfied for $c_1 \cos\phi + c_2 \sin\phi>0$.
Therefore, Eq.~\eqref{sec:bb-a2HDM-3} is satisfied for $c_1 \geq 0$ and $c_2\geq 0$.
In the following, we simplify Eq.~\eqref{sec:bb-a2HDM-3} for $c_1 < 0$ or $c_2 < 0$.

For $c_1 \cos\phi + c_2 \sin\phi < 0$, we can rewrite Eq.~\eqref{sec:bb-a2HDM-3} as
\begin{align}
\sqrt{
 \Biggl(
\lambda_1  \cos^2\phi
+ \lambda_2  \sin^2\phi
+ \tilde{\lambda}_3  \sin(2\phi)
\Biggr) 
\frac{\lambda_a}{2}
}
>
-\left( c_1  \cos\phi  + c_2 \sin\phi \right) 
.
\end{align}
Since the both side are positive, we can square them and find
\begin{align}
\left( \frac{\lambda_a \lambda_1}{2}  -c_1^2  \right)\cos^2\phi
+ \left( \frac{\lambda_a \lambda_2}{2} - c_2^2 \right) \sin^2\phi
+
\left( \frac{\lambda_a \tilde{\lambda}_3}{2}  -c_1 c_2 \right)
\sin2\phi
>
0
.
\label{vtilde-new}
\end{align}

We start from the case for $c_1 > 0$ and $c_2 <0$. In this case, $c_1 \cos\phi + c_2 \sin\phi <0$ for $\phi_0 < \phi < \pi/2$, where $\tan\phi_0 = \frac{c_1}{|c_2|}$.
It is useful to define
\begin{align}
f(x)=
A x^2 + B + 2 D x
,
\label{f}
\end{align}
where
\begin{align}
 A =& \frac{\lambda_a \lambda_1}{2}  -c_1^2,\\
 B =& \frac{\lambda_a \lambda_2}{2} - c_2^2,\\
 D =& \frac{\lambda_a \tilde{\lambda}_3}{2}  -c_1 c_2
.
\end{align}
Eq.~\eqref{vtilde-new} is satisfied if $f(x) > 0$ for $ 0 < x < \frac{|c_2|}{c_1}$.
We find 
\begin{align}
f(0) =& \frac{\lambda_a \lambda_2}{2} - c_2^2
= \left(\sqrt{\frac{\lambda_a \lambda_2}{2}} + c_2\right) \left(\sqrt{\frac{\lambda_a \lambda_2}{2}} - c_2\right)
,\\
f(\cot\phi_0)
 =&
\left(
 \frac{|c_2|}{c_1} \sqrt{\frac{\lambda_a \lambda_1}{2}}
-\sqrt{\frac{\lambda_a \lambda_2}{2}}
\right)^2
 + \lambda_a  \frac{|c_2|}{c_1} 
\left(
 \tilde{\lambda}_3
+ \sqrt{\lambda_1\lambda_2}
\right)
.
\end{align}
These two are always positive thanks to $c_2 < 0$,
Eq.~\eqref{eq:BFB_4}, and Eq.~\eqref{eq:BFB_6}.
Therefore, $f(x) > 0$ at the boundary. It is easy to find that $f(x) > 0$ for $0 < x < \frac{|c_2|}{c_1}$
if one of the following conditions is satisfied,
\begin{align}
           & A \leq 0, \\
\text{or } & A > 0 \text{ \quad and \quad} - \frac{D}{A} \leq 0, \\
\text{or } & A > 0 \text{ \quad and \quad} - \frac{D}{A} \geq \frac{|c_2|}{c_1}, \\
\text{or } & A > 0 \text{ \quad and \quad} 0< - \frac{D}{A} < \frac{|c_2|}{c_1} 
\text{\quad and \quad} B - \frac{D^2}{A} > 0.
\end{align}
The first condition is that $f(x)$ is convex upward.
The second and third conditions are for the $min(f(x))$ is out of $0 < x < \cot\phi_0$.
The last condition is that the minimum exists for $0 < x < \cot\phi_0$ and it is positive.
These conditions are simplified as
\begin{align}
           & A \leq \frac{c_1}{|c_2|}\sqrt{B}, \\
\text{or } & A < \frac{c_1}{|c_2|}\sqrt{B} \text{ \quad and \quad}  -\sqrt{AB} < D ,\\
\text{or } & A < \frac{c_1}{|c_2|}\sqrt{B} \text{ \quad and \quad} D \leq -A\frac{|c_2|}{c_1}.
\label{eq:dame-condition}
\end{align}
After substituting $A$, $B$, and $D$ into these conditions, we find that
$\tilde{V}$ is positive for $c_1 > 0$ and $c_2 < 0$ if
\begin{align}
           & \lambda_1 \leq \lambda_2 \frac{c_1^2}{c_2^2}, \label{eq:cond1}\\
\text{or } & \lambda_1 > \lambda_2 \frac{c_1^2}{c_2^2} 
       \text{ \quad and \quad} 
           \frac{\lambda_a \tilde{\lambda}_3}{2} - c_1 c_2 + \sqrt{\left( \frac{\lambda_a \lambda_1}{2} - c_1^2  \right)\left( \frac{\lambda_a \lambda_2}{2} - c_2^2  \right)} > 0.
\label{eq:cond2}.
\end{align}
We find that Eq.~\eqref{eq:dame-condition} is inconsistent with Eq.~\eqref{eq:BFB_4}.

In a similar manner, we find the conditions for $c_1 < 0$ and $c_2 > 0$ as
\begin{align}
           & \lambda_2 \leq \lambda_1 \frac{c_2^2}{c_1^2},\\
\text{or } & \lambda_2 > \lambda_1 \frac{c_2^2}{c_1^2} 
       \text{ \quad and \quad} 
           \frac{\lambda_a \tilde{\lambda}_3}{2} - c_1 c_2 + \sqrt{\left( \frac{\lambda_a \lambda_1}{2} - c_1^2  \right)\left( \frac{\lambda_a \lambda_2}{2} - c_2^2  \right)} > 0.
.
\end{align}
For $c_1 < 0$ and $c_2 < 0$, $c_1 = 0$ and $c_2 < 0$, or $c_1 < 0$ and $c_2 = 0$, we find
\begin{align}
 \frac{\lambda_a \tilde{\lambda}_3}{2} - c_1 c_2 + \sqrt{\left(\frac{\lambda_a \lambda_1}{2} - c_1^2\right) \left( \frac{\lambda_a \lambda_2}{2} - c_2^2  \right)} > 0.
 \label{eq:cond5}
\end{align}

Eqs.~\eqref{eq:cond1}--\eqref{eq:cond5} are summarized as follows.
\begin{align}
&
\begin{cases}
\lambda_1 \leq \lambda_2 \frac{c_1^2}{c_2^2},\\
\lambda_1 > \lambda_1 \frac{c_1^2}{c_2^2} 
       \text{ \quad and \quad} 
           \frac{\lambda_a \tilde{\lambda}_3}{2} - c_1 c_2 + \sqrt{\left( \frac{\lambda_a \lambda_1}{2} - c_1^2  \right)\left( \frac{\lambda_a \lambda_2}{2} - c_2^2  \right)} > 0.
\end{cases}
&
(c_1 > 0, c_2 < 0)
\\
&
\begin{cases}
\lambda_2 \leq \lambda_1 \frac{c_2^2}{c_1^2},\\
\lambda_2 > \lambda_1 \frac{c_2^2}{c_1^2} 
       \text{ \quad and \quad} 
           \frac{\lambda_a \tilde{\lambda}_3}{2} - c_1 c_2 + \sqrt{\left( \frac{\lambda_a \lambda_1}{2} - c_1^2  \right)\left( \frac{\lambda_a \lambda_2}{2} - c_2^2  \right)} > 0.
\end{cases}
& (c_1 < 0, c_2 > 0)
\\
&
 \frac{\lambda_a \tilde{\lambda}_3}{2} - c_1 c_2 + \sqrt{\left( \frac{\lambda_a \lambda_1}{2} - c_1^2  \right)\left( \frac{\lambda_a \lambda_2}{2} - c_2^2  \right)} > 0.
& (c_1 \leq 0, c_2 \leq 0)
.
\end{align}
They can be further simplified as follows.
\begin{align}
\begin{cases}
\sqrt{\lambda_1} c_2 + \sqrt{\lambda_2} c_1 \geq 0,\\
\sqrt{\lambda_1} c_2 + \sqrt{\lambda_2} c_1 <0
\text{ and } 
\frac{\lambda_a \tilde{\lambda}_3}{2} - c_1 c_2 + \sqrt{\left( \frac{\lambda_a \lambda_1}{2} - c_1^2  \right)\left( \frac{\lambda_a \lambda_2}{2} - c_2^2  \right)} > 0.
\end{cases}
\label{eq:final-condition}
\end{align}
Eq.~\eqref{eq:final-condition} should be satisfied for any $c_1$ and $c_2$.

\section{Beta functions}
\label{sec:beta-function}

\begin{align}
(4\pi)^2 \mu \frac{d \lambda_a}{d\mu}
=&
18 \lambda_a^2  - 4 g_\chi^4  + 4\lambda_a g_\chi^2  + 8 c_1^2 + 8 c_2^2 ,\\
(4\pi)^2 \mu \frac{d c_1}{d\mu}
=&
 8 c_1^2 - \frac{3}{2} c_1 \left( g_1^2 + 3 g_2^2 \right)
 + 2 c_1 \left( 3 \lambda_1 + 3 \lambda_a + g_\chi^2 \right)
 + 2 c_2 \left( 2 \lambda_3 + \lambda_4 \right)
,\\
(4\pi)^2 \mu \frac{d c_2}{d\mu}
=&
 8 c_2^2 - \frac{3}{2} c_2 \left(g_1^2 + 3 g_2^2 \right)
 + 2 c_2 \left( 3 \lambda_2 + 3 \lambda_a + g_\chi^2 +3 y_t^2 \right)
 + 2 c_1 \left( 2 \lambda_3 + \lambda_4 \right)
,\\
(4\pi)^2 \mu \frac{d g_\chi}{d\mu}
=&
 4 g_\chi^3.
\end{align}

\end{document}